\documentclass[conference]{IEEEtran}

\usepackage{standalone}

\usepackage{graphicx}
\usepackage{pdfpages}

\usepackage{caption}

\usepackage{colortbl}
\usepackage{xcolor}
\usepackage{array}
\usepackage{makecell}

\usepackage{soul}
\sethlcolor{yellow}

\usepackage{tabularx}

\newcommand{\HiLi}[1]{%
  \leavevmode%
  \rlap{%
    \hspace*{-2pt}%
    \hbox to \dimexpr\hsize + #1\relax{%
      \color{gray!20}%
      \leaders\hrule height .8\baselineskip depth .8ex\hfill%
    }%
  }%
}

\newcommand{\HiliColor}[2]{
	\leavevmode%
	\rlap{%
	  \hspace*{-1pt}%
	  \hbox to \dimexpr\hsize + #2\relax{%
		\color{#1}%
		\leaders\hrule height .8\baselineskip depth .8ex\hfill%
	  }%
	}%
}

\usepackage{tikz}
\usetikzlibrary{fit,calc}

\colorlet{mypink}{red!40}
\colorlet{myblue}{cyan!40}
\colorlet{mygray}{gray!20}
\colorlet{myyellow}{yellow!40}

\usetikzlibrary{tikzmark}

\let\oldnl\nl%
\newcommand{\nonl}{\renewcommand{\nl}{\let\nl\oldnl}}%

\usepackage{mdframed} %

\definecolor{light}{RGB}{222,235,247}
\definecolor{dark}{RGB}{158,202,225}
\definecolor{darker}{RGB}{49,130,189}
\newcolumntype{a}{>{\columncolor{light}}r}
\newcolumntype{b}{>{\columncolor{dark}}r}
\usepackage[ruled,lined,linesnumbered,vlined]{algorithm2e}

\SetCommentSty{mycommfont}

\usepackage{adjustbox}

\usepackage{amsmath}
\usepackage{amsfonts}
\usepackage{amsopn}
\usepackage{amsthm}
\usepackage{nccmath} %
\usepackage{url} %
\usepackage{multicol}
\usepackage{multirow}
\usepackage{setspace} %
\usepackage{xspace}
\usepackage{listings}
\usepackage{stmaryrd} %
\usepackage{todonotes} %
\usepackage{booktabs}
\usepackage{threeparttable}
\usepackage[utf8x]{inputenc}
\usepackage[normalem]{ulem}	%
\usepackage{balance}
\usepackage[labelformat=simple]{subcaption}

\usepackage{tcolorbox}

\usepackage[%
  pdfborder={0 0 0}, %
]{hyperref}

\usepackage[scaled=0.87]{beramono}

\usepackage{enumitem}
\usepackage{pdftexcmds}
\usepackage{microtype}
\usepackage{flushend}
\usepackage{pgf}

\newcommand{\proj}{\textsf{WeightDD}\xspace}

\newcommand{\testInput}{\ensuremath{\textit{l}}\xspace}
\newcommand{\prefix}{\ensuremath{\textit{pre}}\xspace}
\newcommand{\testInputMin}{\ensuremath{\testInput{}_\textit{min}}\xspace}

\newcommand{\sortedTestInput}{\ensuremath{l_\textit{sorted}}\xspace}

\newcommand{\wdd}{WDD\xspace}
\newcommand{\wprobdd}{\ensuremath{\mbox{W}_\text{ProbDD}}\xspace}
\newcommand{\wddmin}{\ensuremath{\mbox{W}_\text{ddmin}}\xspace}
\newcommand{\ddmin}{ddmin\xspace}
\newcommand{\dd}{DD\xspace}
\newcommand{\hdd}{HDD\xspace}
\newcommand{\probdd}{ProbDD\xspace}
\newcommand{\perses}{Perses\xspace}
\newcommand{\deltadebugging}{Delta Debugging\xspace}
\newcommand{\minimizingdeltadebugging}{Minimizing Delta Debugging\xspace}
\newcommand{\hierarchicaldeltadebugging}{Hierarchical Delta Debugging\xspace}
\newcommand{\weighteddeltadebugging}{Weighted Delta Debugging\xspace}
\newcommand{\weightedMinimizingDeltaDebugging}{Weighted Minimizing Delta Debugging\xspace}
\newcommand{\probabilisticdeltadebugging}{Probabilistic Delta Debugging\xspace}
\newcommand{\weightedprobabilisticdeltadebugging}{Weighted Probabilistic Delta Debugging\xspace}

\newcommand{\emptylist}{\ensuremath{[\hspace{0.5\fontdimen2\font}]}}

\newcommand{\property}{\ensuremath{\psi}\xspace}

\newcommand{\weight}{\ensuremath{w}\xspace}
\newcommand{\weightFullTerm}{\ensuremath{\textit{weight}}\xspace}
\newcommand{\partitions}{\ensuremath{\textit{partitions}}\xspace}

\newcommand{\partition}{\ensuremath{\textit{ptn}}\xspace}

\newcommand{\p}{\ensuremath{\textit{p}}}
\newcommand{\e}{\ensuremath{\textit{e}}}
\newcommand{\pOne}{\ensuremath{p_1}\xspace}
\newcommand{\pTwo}{\ensuremath{p_2}\xspace}
\newcommand{\pinit}{\ensuremath{p_0}\xspace}

\newcommand{\probs}{\ensuremath{\textit{probs}}\xspace}

\newcommand{\node}{\ensuremath{\textit{element}}\xspace}

\newcommand{\gain}{\ensuremath{\textit{gain}}\xspace}

\newcommand{\probOfDeletion}{\ensuremath{\textit{probOfDeletion}}\xspace}

\newcommand{\pDel}{\ensuremath{\textit{P}_\textit{del}}\xspace}

\newcommand{\tokensPerSecond}{\ensuremath{\textit{tokens/s}}\xspace}

\SetKwFunction{FuncWDD}{wddRec}
\SetKwFunction{FuncWeightPartition}{weightedPartition}
\SetKwFunction{FuncCheckOneMinimal}{ensureOneMinimal}

\SetKwFunction{FuncGetWeight}{getWeight}
\SetKwFunction{FuncInitProbability}{initProbs}
\SetKwFunction{FuncShouldTerminate}{shouldTerminate}
\SetKwFunction{FuncSort}{sort}
\SetKwFunction{FuncGetNextTest}{getNextTest}
\SetKwFunction{FuncGetNodesToRemove}{getPartitionToRemove}
\SetKwFunction{FuncUpdateProbability}{updateProbs}

\SetKwData{AlgL}{L}
\SetKwData{AlgP}{P}
\SetKwData{AlgW}{w}

\SetKwData{AlgPartition}{partition}
\SetKwData{AlgInputListMin}{L\_min}
\SetKwData{AlgP}{p}
\SetKwData{AlgPOne}{p1}
\SetKwData{AlgPTwo}{p2}

\SetKwData{AlgNode}{node}
\SetKwData{AlgPartitions}{partitions}
\SetKwData{AlgComplements}{complements}
\SetKwData{AlgComplement}{complement}
\SetKwData{AlgSingleNodeList}{singleNodeList}
\SetKwData{AlgNewPartitions}{result}
\SetKwData{AlgTrue}{true}
\SetKwData{AlgFalse}{false}
\SetKwData{AlgWeight}{half\_weight}
\SetKwData{AlgCurrentWeight}{cur\_weight}
\SetKwData{AlgProbs}{probs}
\SetKwData{AlgProb}{prob}
\SetKwData{AlgWeight}{weight}
\SetKwData{AlgToBeDeleted}{toBeDeleted}
\SetKwData{AlgNextTest}{nextTest}
\SetKwData{AlgTmp}{tmp}
\SetKwData{AlgProduct}{product}
\SetKwData{AlgGain}{gain}
\SetKwData{AlgLastGain}{lastGain}
\SetKwData{AlgMaxGain}{maxGain}

\newcommand{\complement}{\ensuremath{\textit{complement}}\xspace}
\newcommand{\maxGain}{\ensuremath{\textit{gain}_\textit{max}}\xspace}
\newcommand{\halfWeight}{\ensuremath{\textit{halfSum}}\xspace}
\newcommand{\newPartitions}{\ensuremath{\textit{result}}\xspace}

\newcommand{\ddminQueriesOnExample}{30\xspace}
\newcommand{\wddQueriesOnExample}{26\xspace}
\newcommand{\probddQueriesOnExample}{15\xspace}
\newcommand{\wprobddQueriesOnExample}{11\xspace}

\newcommand{\nodeOneTokens}{5\xspace}
\newcommand{\nodeSevenTokens}{25\xspace}
\newcommand{\ddminExampleStdDev}{7.43\xspace}
\newcommand{\wddExampleStdDev}{6.34\xspace}

\newcommand{\cProgramsAverageTokens}{77,723\xspace}
\newcommand{\xmlProgramsAverageTokens}{20,197\xspace}

\newcommand{\hddDD}{HDD$_\text{d}$\xspace}
\newcommand{\hddProbDD}{HDD$_\text{p}$\xspace}
\newcommand{\hddWDD}{HDD$_\text{w}$\xspace}
\newcommand{\hddWProbDD}{HDD$_\text{wp}$\xspace}

\newcommand{\persesDD}{Perses$_\text{d}$\xspace}
\newcommand{\persesProbDD}{Perses$_\text{p}$\xspace}
\newcommand{\persesWDD}{Perses$_\text{w}$\xspace}
\newcommand{\persesWProbDD}{Perses$_\text{wp}$\xspace}

\newcommand{\cBenchmarkSize}{32\xspace}

\newcommand{\xmlBenchmarkSize}{30\xspace}
\newcommand{\allBenchmarksSize}{62\xspace}

\newcommand{\hddWddAverageTime}{51.31\%\xspace}
\newcommand{\hddWddAverageSize}{9.12\%\xspace}
\newcommand{\persesWddAverageTime}{7.47\%\xspace}
\newcommand{\persesWddAverageSize}{0.96\%\xspace}

\newcommand{\hddWprobddAverageTime}{11.98\%\xspace}
\newcommand{\hddWprobddAverageSize}{13.40\%\xspace}
\newcommand{\persesWprobddAverageTime}{9.72\%\xspace}
\newcommand{\persesWprobddAverageSize}{2.20\%\xspace}

\usepackage{siunitx}
\newcommand{\pValHddWddTime}{\num{5.3e-11}\xspace}
\newcommand{\pValHddWddSize}{\num{5.06e-5}\xspace}
\newcommand{\pValPersesWddTime}{\num{2.33e-6}\xspace}
\newcommand{\pValPersesWddSize}{\num{0.53}\xspace} %

\newcommand{\pValHddWprobddTime}{\num{1.06e-6}\xspace}
\newcommand{\pValHddWprobddSize}{\num{1.10e-6}\xspace}
\newcommand{\pValPersesWprobddTime}{\num{1.73e-5}\xspace}
\newcommand{\pValPersesWprobddSize}{\num{0.42}\xspace} %

\newcommand{\spearmanMeanOfHDDDdminOnC}{-0.41\xspace}
\newcommand{\spearmanMeanOfPersesDdminOnC}{-0.14\xspace}
\newcommand{\spearmanMeanOfHDDDdminOnXml}{-0.83\xspace}
\newcommand{\spearmanMeanOfPersesDdminOnXml}{-0.36\xspace}

\newcommand{\hddWddTimeOnC}{53.92\%\xspace}
\newcommand{\hddWddTimeOnXML}{16.86\%\xspace}
\newcommand{\persesWddTimeOnC}{9.01\%\xspace}
\newcommand{\persesWddTimeOnXML}{1.67\%\xspace}

\newcommand{\hddWddSizeOnC}{7.81\%\xspace}
\newcommand{\hddWddSizeOnXML}{16.51\%\xspace}
\newcommand{\persesWddSizeOnC}{1.04\%\xspace}
\newcommand{\persesWddSizeOnXML}{0.27\%\xspace}

\newcommand{\hddWddAverageTokensPerSecond}{14.2\xspace}
\newcommand{\hddDdminAverageTokensPerSecond}{6.59\xspace}
\newcommand{\hddWddTokensPerSecondRatio}{115.57\%\xspace}

\newcommand{\persesWddAverageTokensPerSecond}{40.2\xspace}
\newcommand{\persesDdminAverageTokensPerSecond}{38.14\xspace}
\newcommand{\persesWddTokensPerSecondRatio}{5.40\%\xspace}

\newcommand{\hddWprobddTimeOnC}{10.89\%\xspace}
\newcommand{\hddWprobddTimeOnXML}{14.31\%\xspace}
\newcommand{\persesWprobddTimeOnC}{13.89\%\xspace}
\newcommand{\persesWprobddTimeOnXML}{1.35\%\xspace}

\newcommand{\hddWprobddSizeOnC}{13.91\%\xspace}
\newcommand{\hddWprobddSizeOnXML}{9.74\%\xspace}
\newcommand{\persesWprobddSizeOnC}{2.43\%\xspace}
\newcommand{\persesWprobddSizeOnXML}{0.32\%\xspace} %

\newcommand{\persesWprobddSameResultCount}{36\xspace}

\newcommand{\persesWprobddSameResultCountOnXML}{29\xspace}

\newcommand{\hddWprobddAverageTokensPerSecond}{24.03\xspace}
\newcommand{\hddProbddAverageTokensPerSecond}{15.82\xspace}
\newcommand{\hddWprobddTokensPerSecondRatio}{51.90\%\xspace}
\newcommand{\persesWprobddAverageTokensPerSecond}{40.36\xspace}
\newcommand{\persesProbddAverageTokensPerSecond}{39.95\xspace}
\newcommand{\persesWprobddTokensPerSecondRatio}{1.03\%\xspace}

\makeatletter

\newcommand{\Rmnum}[1]{\expandafter\@slowromancap\romannumeral
#1@}
\makeatother

\newcommand{\gcc}{GCC\xspace}
\newcommand{\llvm}{LLVM\xspace}

\newcommand{\creduce}{C-Reduce\xspace}
\newcommand{\jreduce}{J-Reduce\xspace}

\newcommand{\jsdelta}{JS Delta\xspace}
\newcommand{\ddsmt}{ddSMT\xspace}

\newcommand{\vulcan}{Vulcan\xspace}

\newcommand{\searchspace}{\ensuremath{\mathbb{L}}\xspace}

\newcommand{\boolspace}{\ensuremath{\mathbb{B}}\xspace}
\newcommand{\naturalnumbers}{\ensuremath{\mathbb{N}}\xspace}
\newcommand{\elementspace}{\ensuremath{\mathbb{E}}\xspace}

\newcommand{\aka}{\hbox{\emph{a.k.a.}}\xspace}

\newcommand{\etal}{\hbox{\emph{et al.}}\xspace}
\newcommand{\vs}{\hbox{\emph{v.s.}}\xspace}
\newcommand{\eg}{\hbox{\emph{e.g.}}\xspace}
\newcommand{\ie}{\hbox{\emph{i.e.}}\xspace}

\SetKwFunction{AlgReduce}{Reduce}
\SetKwFunction{AlgRemove}{Remove}
\SetKwFunction{AlgMinSubtree}{MinSubtree}
\SetKwFunction{AlgUndoReplace}{UndoReplace}
\SetKwFunction{AlgInsert}{Insert}
\SetKwFunction{AlgDelete}{Delete}
\SetKwFunction{AlgReplace}{Replace}
\SetKwFunction{AlgMain}{\proj}
\SetKwFunction{AlgPatch}{Patch}
\SetKwFunction{AlgTreeDiff}{$\Delta_T$}
\SetKwFunction{AlgListDiff}{$\Delta_L$}
\SetKwFunction{AlgPop}{Pop}
\SetKwFunction{AlgPush}{Push}
\SetKwFunction{AlgRemainingNodes}{GetRemainingNodes}
\SetKwFunction{AlgReducePlus}{ReducePlus}
\SetKwFunction{AlgReduceOptional}{ReduceOptional}
\SetKwFunction{AlgReduceRegular}{ReduceRegular}
\SetKwFunction{AlgGetAndRemoveLargestFrom}{GetAndRemoveLargestFrom}
\SetKwFunction{AlgChildren}{Children}
\SetKwFunction{AlgParent}{Parent}
\SetKwFunction{AlgDdmin}{ddmin}
\SetKwFunction{AlgDd}{dd}
\SetKwFunction{AlgParse}{ParseTree}
\SetKwFunction{AlgCopy}{Duplicate}
\SetKwFunction{AlgGetTransformationList}{getTransformList}
\SetKwFunction{AlgGetTargetList}{getTargetList}
\SetKwFunction{AlgApplyTransformation}{applyTransformation}
\SetKwFunction{AlgGetPrimaryQuestion}{getPrimaryQuestion}
\SetKwFunction{AlgGetFollowupQuestion}{getFollowupQuestion}
\SetKwFunction{AlgPerses}{Perses}

\SetKwData{AlgMin}{Min}
\SetKwData{AlgMinIndex}{min\_index}
\SetKwData{AlgResult}{result}

\SetKwFunction{Encode}{Encode}
\SetKwFunction{CompactEncode}{CompactEncode}
\SetKwFunction{Decode}{CompactDecode}
\SetKwFunction{Refresh}{RefreshCache}

\SetKwFunction{BoundedBFS}{BoundedBFS}
\SetKwFunction{IsKleene}{IsKleene}
\SetKwFunction{RuleType}{Rule}
\SetKwFunction{ExpectedRuleType}{ExpectedRule}
\SetKwFunction{QuantifiedRuleType}{QuantifiedRule}

\SetKwData{Candidates}{candidates}
\SetKwData{Predicate}{pred}

\SetKwData{AlgTrueLiteral}{\mycode{T}}
\SetKwData{AlgFalseLiteral}{\mycode{F}}
\SetKwData{AlgMin}{min}
\SetKwData{AlgCache}{cache}
\SetKwData{AlgPrev}{prev}
\SetKwData{AlgWorklist}{worklist}
\SetKwData{AlgNext}{pending}
\SetKwData{AlgLargest}{largest}

\SetKw{Continue}{continue}

\SetKwProg{Fn}{Function}{:}{}

\newtheorem{definition}{Definition}[section]

\newtheorem{Assumption}{Assumption}[section]

\newcommand{\mycode}[1]{\texttt{#1}\xspace}

\newcommand{\HighlightBlack}[1]{}

\newcommand{\myparagraph}[1]{
	\vspace*{0.04cm}
	\noindent \textit{\textbf{#1.}}\xspace\xspace
}

\usepackage{tikz}
\newcommand*\circled[1]{\tikz[baseline=(char.base)]{
		\node[shape=circle,draw,inner sep=1pt] (char) {#1};}}

\usepackage{tikz}
\usepackage{forest}

\definecolor{deepRed}{HTML}{D81B60}
\definecolor{deepBlue}{HTML}{1E88E5}
\definecolor{deepOrange}{HTML}{FFC107}

\newcounter{num}

\newcommand{\finding}[1]{
	\begin{tcolorbox}[arc=2pt,left=2pt,right=2pt,top=2pt,bottom=2pt,boxsep=0pt]
		\textbf{RQ\refstepcounter{num}\thenum}: #1
	\end{tcolorbox}
}

\usepackage{cleveref} %

\Crefname{table}{Table}{Tables}
\crefname{table}{Table}{Tables}
\Crefname{figure}{Fig.}{Figs.}
\crefname{figure}{Fig.}{Figs.}
\Crefname{algocf}{Algorithm}{Algorithms}
\crefname{algocf}{Algorithm}{Algorithms}
\Crefname{algorithm}{Algorithm}{Algorithms}
\crefname{algorithm}{Algorithm}{Algorithms}
\crefname{thm}{Theorem}{Theorems}
\Crefname{thm}{Theorem}{Theorems}
\Crefname{Assumption}{Assumption}{Assumptions}
\crefname{appendix}{Appendix}{Appendices}
\Crefname{appendix}{Appendix}{Appendices}

\crefformat{chapter}{\S#2#1#3}
\crefmultiformat{chapter}{\S\S#2#1#3}{ and~#2#1#3}{, #2#1#3}{, and~#2#1#3}

\crefformat{section}{\S#2#1#3}
\crefmultiformat{section}{\S\S#2#1#3}{ and~#2#1#3}{, #2#1#3}{, and~#2#1#3}

\def\BibTeX{{\rm B\kern-.05em{\sc i\kern-.025em b}\kern-.08em
        T\kern-.1667em\lower.7ex\hbox{E}\kern-.125emX}}

\begin{document}

\title{\wdd: Weighted Delta Debugging}

\makeatletter
\newcommand{\linebreakand}{%
  \end{@IEEEauthorhalign}
  \hfill\mbox{}\par
  \mbox{}\hfill\begin{@IEEEauthorhalign}
}
\makeatother

\author{
	\IEEEauthorblockN{
    Xintong Zhou\IEEEauthorrefmark{1},
    Zhenyang Xu\IEEEauthorrefmark{1},
		Mengxiao Zhang\IEEEauthorrefmark{1},
		Yongqiang Tian\IEEEauthorrefmark{2},
		and Chengnian Sun\IEEEauthorrefmark{1}
	}
	\IEEEauthorblockA{
		\IEEEauthorrefmark{1}School of Computer Science, University of
		Waterloo, Waterloo, Canada\\
		Emails: x27zhou@uwaterloo.ca, zhenyang.xu@uwaterloo.ca, m492zhan@uwaterloo.ca, cnsun@uwaterloo.ca
	}
	\IEEEauthorblockA{
		\IEEEauthorrefmark{2}Department of Computer Science and
		Engineering,\\
		The Hong Kong University of Science and Technology, Hong Kong,
		China\\
		Email: yqtian@ust.hk
	}
}

\maketitle

\begin{abstract}
	Delta Debugging is a widely used family of algorithms (\eg, \ddmin and \probdd)
to automatically minimize bug-triggering test inputs, thus to facilitate debugging.
It takes a list of elements with each element representing a fragment of the test input,
systematically partitions the list at different granularities,
identifies and deletes bug-irrelevant partitions.

Prior delta debugging algorithms assume there are no differences among the elements in the list,
and thus treat them uniformly during partitioning.
However, in practice, this assumption usually does not hold,
because the size (referred to as weight) of the fragment
represented by each element can vary significantly.
For example, a single element representing 50\% of the test input
is much more likely to be bug-relevant than elements representing only 1\%.
This assumption inevitably impairs the efficiency
or even effectiveness of these delta debugging algorithms.

This paper proposes \weighteddeltadebugging (\wdd),
a novel concept to help prior delta debugging algorithms
overcome the limitation mentioned above.
The key insight of \wdd is to assign each element in the list a weight according to its size,
and distinguish different elements based on their weights during partitioning.
We designed two new minimization algorithms,
\wddmin and \wprobdd, by applying \wdd to \ddmin and \probdd  respectively.
We extensively evaluated \wddmin and \wprobdd in two representative applications,
\hdd and \perses, on \allBenchmarksSize benchmarks across two languages.
On average, with \wddmin,
\hdd and \perses took \hddWddAverageTime and \persesWddAverageTime less time to
generate \hddWddAverageSize and \persesWddAverageSize smaller results than with \ddmin, respectively.
With \wprobdd, \hdd and \perses used \hddWprobddAverageTime and \persesWprobddAverageTime
less time to generate \hddWprobddAverageSize and \persesWprobddAverageSize smaller results than with \probdd, respectively.
The results strongly demonstrate the value of \wdd.
We firmly believe that \wdd opens up a new dimension to improve test input minimization techniques.

\end{abstract}


\section{Introduction}
\label{sec:intro}
A bug-triggering test input, which causes a program to fail,
often contains many bug-irrelevant elements.
These elements usually complicate the
use of the test input to debug the program.
Test input minimization is a
technique that automatically minimizes the size of the input by
removing the irrelevant elements while keeping the failure-inducing parts.
It helps developers to focus on the essential parts of the input that cause the failure.
Many minimization techniques~\cite{zeller2002simplifying,xu2024t,xu2023pushing,rcc,
misherghi2006hdd, regehr2012test,sun2018perses,adhoc,zhang2023ppr,zhang2024lpr}
have been proposed and widely used in various scenarios~\cite{chisel,chen2013taming,
rubio2013precimonious,reducersarefuzzers,binkley2014orbs},
especially in facilitating software testing and debugging~\cite{chen2020survey,yang2011finding,ccmd}.

Delta Debugging~\cite{zeller2002simplifying} is a widely used family of algorithms
to automatically minimize bug-triggering test inputs.
Typically, delta debugging algorithms
take a
test input as a list of elements,
with each element representing a fragment of the test input (\eg, a token, a line, or a tree node).
Then it partitions the list into sets of elements (referred to as partitions) at different granularities,
systematically identifies and deletes partitions that are bug-irrelevant.
State-of-the-art algorithms in this family include
\minimizingdeltadebugging (\ddmin)~\cite{zeller2002simplifying} and
\probabilisticdeltadebugging (\probdd)~\cite{wang2021probabilistic}.
The first delta debugging algorithm \ddmin
systematically minimizes the list of elements in a binary-search style.
The generality, effectiveness, and efficiency of \ddmin
make it a fundamental minimization algorithm in many subsequently proposed minimization tools~\cite{misherghi2006hdd,sun2018perses}.
The other algorithm, \probdd~\cite{wang2021probabilistic}
is a recently proposed variant of \ddmin.
It improves the efficiency of \ddmin by leveraging a probabilistic model
to guide the minimization process.

In practice, delta debugging algorithms are often
applied to the tree representations of the inputs
rather than plain lists of tokens or lines to achieve better minimization performance, \aka, tree-based minimization.
For example, \hierarchicaldeltadebugging (\hdd)
proposed by Misherghi and Su~\cite{misherghi2006hdd}
represents the input as a tree structure (\eg, a parse tree),
and then uses \ddmin to minimize each level of the tree from coarse to fine.
Another example is \perses~\cite{sun2018perses}, a minimization technique
that further improves \hdd by leveraging context-free grammar
to ensure the syntactic validity during minimization.
\perses applies \ddmin on the child node list of quantified nodes
(\ie, a type of nodes whose children are independent to each other in terms of syntax validity)
in the parse tree.
Both \hdd and \perses show significant superiority in handling structured inputs compared to directly
applying delta debugging to the flat list representations of the inputs.

\myparagraph{Limitations}
One significant limitation
of prior delta debugging algorithms~\cite{zeller2002simplifying,wang2021probabilistic}
is that they overlooked
the effect of element size in minimization,
and thus the efficiency or even effectiveness of minimization
is impaired.
Specifically,
\ddmin performs a binary-search style
deletion and iteratively divides the list into smaller partitions
evenly by length (\ie, the number of elements).
However, due to the varying sizes of elements, \ddmin fails to achieve the
true evenness\footnote{
    The true evenness indicates that the size of each partition
    approximately equals to each other. The size of a partition
    is normally measured by the number of tokens it contains,
    that is to say, the number of tokens in each partition is
    approximately equal.
}
and generates partitions with significantly different sizes.
For example, when \hdd invokes \ddmin to minimize
the bug-triggering input of LLVM-19595~\cite{llvm19595report},
the largest and smallest partitions produced by a partitioning
operation can contain 8,752 and 5 tokens, respectively.
However, \ddmin treats these uneven partitions equally,
neglecting an important statistical observation,
\ie, \textit{larger partitions are more likely
to contain the failure-inducing elements and thus less likely
to be removed}.
As a result,
\ddmin spends significant efforts
in removing large but unlikely to be removed elements
during the minimization process,
which restricts its performance in large and complex
bug-triggering inputs.
As for \probdd,
while it successfully refines the partitioning
strategy of \ddmin with its probabilistic model,
it still lacks awareness of the varying sizes of elements during partitioning,
thus leading to suboptimal performance.
More details of this limitation and its affect is illustrated
in \cref{sec:motivation}.

\myparagraph{Weighted Delta Debugging}
In this paper, we propose \weighteddeltadebugging (\wdd),
a novel concept to improve prior delta debugging algorithms
by overcoming the aforementioned limitation.
The key insight of \wdd is to take the sizes of elements into consideration
and assign each element a weight based on its size.
By doing so,
\wdd can perform a more rational weight-based partitioning strategy,
thereby enhancing minimization performance.
We apply \wdd to two representative delta debugging
algorithms, \ddmin and \probdd, and propose two new algorithms,
\wddmin and \wprobdd, respectively.
At a high level,
\wddmin improves \ddmin by performing a weighted
binary-search style minimization, while \wprobdd enhances \probdd
by incorporating the weights of elements as a new factor into the probabilistic model
which guides the partitioning.

We extensively evaluate \wddmin and \wprobdd on \allBenchmarksSize
benchmarks across two languages, \ie, C and XML, by substituting them
for \ddmin and \probdd, respectively, in two application scenarios,
\hdd~\cite{misherghi2006hdd} and \perses~\cite{sun2018perses}.
The results demonstrate that
\wddmin and \wprobdd significantly outperform \ddmin and \probdd
in efficiency and effectiveness, respectively. On average,
after substituting \wddmin for \ddmin, \hdd and \perses use
\hddWddAverageTime and \persesWddAverageTime less time to
produce \hddWddAverageSize and \persesWddAverageSize smaller results, respectively.
Moreover, with \wprobdd, \hdd and \perses obtain
\hddWprobddAverageSize and \persesWprobddAverageSize smaller results
with \hddWprobddAverageTime and \persesWprobddAverageTime less time
than using \probdd,	respectively.

\myparagraph{Contribution}
This paper makes the following contributions.
\begin{itemize}[leftmargin=*]
	\item We present \weighteddeltadebugging (\wdd),
	a novel concept that helps prior delta debugging algorithms
	overcome the limitation of being unaware of the different sizes
	among the elements in the input list.

	\item We realize \wdd in two representative delta debugging
	algorithms, \ddmin and \probdd, and propose two new algorithms,
	\wddmin and \wprobdd, respectively.

	\item We comprehensively evaluate \wddmin and \wprobdd on
	\allBenchmarksSize benchmarks in different application scenarios.
	The results demonstrate the superiority of \wddmin
	and \wprobdd over \ddmin and \probdd, respectively, thus highlighting
	the significance of \wdd in improving test input minimization.

	\item For replication, we make the artifacts of this paper publicly available~\cite{wddartifact}.
	We also release the source code of \wddmin and \wprobdd in the Perses~\cite{persesRepo}
	repository for further research and applications.
\end{itemize}

\section{Background}
\label{sec:background}
Test input minimization facilitates the software debugging process
by automatically minimizing the size of the bug-triggering test input. This technique
is highly demanded as it helps developers to
focus on the essential parts of the test input and saves the time
and effort required to identify the root cause of the bug.
For example, both GCC~\cite{gccReductionGuide} and LLVM~\cite{llvmReductionGuide}
have explicitly announced that the bug-triggering program should
be minimized before being reported.
Test input minimization also assists many other software engineering tasks,
such as program analysis~\cite{rubio2013precimonious}
and slicing~\cite{binkley2014orbs}.

To facilitate presentation, we introduce the notations below,
\begin{itemize}[topsep=0pt, leftmargin=*]
\item $\elementspace$ denotes the set of all possible elements in  test inputs

\item $\testInput$ denotes a test input, which is a list of elements with elements drawn from \elementspace

\item $\searchspace$ denotes the universe of possible test inputs, namely, $\testInput \in \searchspace$.

\item $\boolspace=\{\AlgTrueLiteral, \AlgFalseLiteral\}$ where \AlgTrueLiteral for true and \AlgFalseLiteral for false.

\item $\property:\searchspace\rightarrow\boolspace$ is a property test function returning \AlgTrueLiteral if
the given input preserves a certain property, \AlgFalseLiteral otherwise.

\item $\weight:\elementspace\rightarrow \naturalnumbers$ is a weight function computing the weight (a natural number, such as 0, 1, and 2) of  an element.
\end{itemize}
With these symbols, the problem of test input minimization can be formalized as follows.
\begin{definition}[Test Input Minimization]
Given a test input $\testInput\in\searchspace$
for a program
and a property \property
exhibited by \testInput, e.g., triggering a bug or generating an
unexpected output when the program executes with \testInput,
the objective of test input minimization is to produce a
 test input $\testInputMin\in\searchspace$ that has a minimal number of elements and
 still exhibits \property, i.e.,
$\property(\testInputMin)=\AlgTrueLiteral$.
\end{definition}

\begin{figure*}[ht]
    \centering
    \begin{subfigure}[b]{0.34\textwidth}
        \centering

\begin{lstlisting}[firstnumber=1, numbers=left, numbersep=4pt]
typedef long long llong; @\textcolor{orange}{...........\circled{1} $\weight= 5$}@
test2char64(char *p) {} @\textcolor{orange}{............\circled{2} $\weight= 8$}@
test1char8(char c) {} @\textcolor{orange}{..............\circled{3} $\weight= 7$}@
test1short32(short c) {} @\textcolor{orange}{...........\circled{4} $\weight= 7$}@
test2short32(short *p) {} @\textcolor{orange}{..........\circled{5} $\weight= 8$}@
typedef llong vllong1 \
\end{lstlisting}

\begin{lstlisting}[]
__attribute__(( \
__vector_size__(sizeof(llong)))); @\textcolor{orange}{..\circled{6} $\weight= 16$}@
\end{lstlisting}

\begin{lstlisting}[firstnumber=7, numbers=left, numbersep=4pt,]
vllong1 test2llong1(llong *p) {
\end{lstlisting}

\begin{lstlisting}
    llong c = *test1char8;
    vllong1 v = {c};
    return v;
} @\textcolor{orange}{..................................\circled{7} $\weight= 25$}@
\end{lstlisting}

\begin{lstlisting}[firstnumber=8, numbers=left, numbersep=4pt,]
int main() {} @\textcolor{orange}{......................\circled{8} $\weight= 6$}@
\end{lstlisting}

       \caption{A program that triggers \gcc to crash.
       }
       \label{subfig:example}
    \end{subfigure}
        \hfil
    \begin{subfigure}[b]{0.3\textwidth}
        \begin{tikzpicture}[
            every node/.style={draw, rectangle, rounded corners, align=center, line width=0.4pt, font=\scriptsize},
            level distance=1.2cm,
            level 1/.style={sibling distance=2.7cm},
            level 2/.style={sibling distance=1.45cm},
            level 3/.style={sibling distance=0.65cm}
        ]
            \node {1--8\textcolor{orange}{:82}}
              child {node {1--4\textcolor{orange}{:27}}
                child {node {1,2\textcolor{orange}{:13}}
                    child {node {1\textcolor{orange}{:5}}}
                    child {node {2\textcolor{orange}{:8}}}
                }
                child {node {3,4\textcolor{orange}{:14}}
                    child {node {3\textcolor{orange}{:7}}}
                    child {node {4\textcolor{orange}{:7}}}
                }
              }
              child {node {5--8\textcolor{orange}{:55}}
                child {node {5,6\textcolor{orange}{:24}}
                    child {node {5\textcolor{orange}{:8}}}
                    child {node {6\textcolor{orange}{:16}}}
                }
                child {node {7,8\textcolor{orange}{:31}}
                    child {node {7\textcolor{orange}{:25}}}
                    child {node {8\textcolor{orange}{:6}}}
                }
              };
          \end{tikzpicture}
        \caption{
        	The search space of \ddmin.
        }
        \label{subfig:dd-searchspace}
    \end{subfigure}
        \hfil
    \begin{subfigure}[b]{0.3\textwidth}
        \begin{tikzpicture}[
            every node/.style={draw, rectangle, rounded corners, align=center, line width=0.4pt, font=\scriptsize},
            level distance=0.9cm,
            level 1/.style={sibling distance=2.5cm},
            level 2/.style={sibling distance=1.4cm},
            level 3/.style={sibling distance=0.75cm},
            level 4/.style={sibling distance=0.7cm}
        ]
            \node {1--8\textcolor{orange}{:82}}
              child{node {1--5\textcolor{orange}{:35}}
                child {node {1--3\textcolor{orange}{:20}}
                    child {node {1,2\textcolor{orange}{:13}}
                        child {node {1\textcolor{orange}{:5}}}
                        child {node {2\textcolor{orange}{:8}}}
                    }
                    child {node {3\textcolor{orange}{:7}}}
                }
                child {node {4,5\textcolor{orange}{:15}}
                    child {node {4\textcolor{orange}{:7}}}
                    child {node {5\textcolor{orange}{:8}}}
                }
              }
              child {node {6--8\textcolor{orange}{:47}}
                child {node {6\textcolor{orange}{:16}}}
                child {node {7,8\textcolor{orange}{:31}}
                    child {node {7\textcolor{orange}{:25}}}
                    child {node {8\textcolor{orange}{:6}}}
                }
              };
          \end{tikzpicture}
        \caption{The search space of \wddmin.}
        \label{subfig:wdd-searchspace}
    \end{subfigure}

    \caption{A motivating example. In each subfigure, the weights of the nodes or the partitions are highlighted in orange.}
    \label{fig:motivating_example}

\end{figure*}
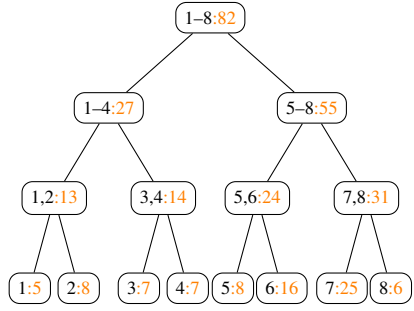
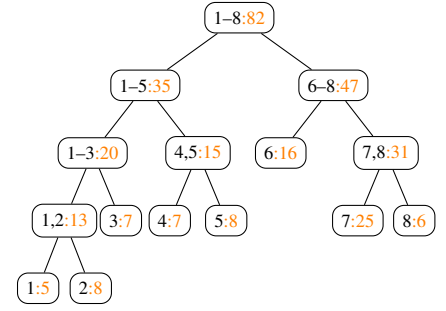

Many techniques~\cite{zeller2002simplifying,misherghi2006hdd,regehr2012test,sun2018perses,wang2021probabilistic,zhang2025toward,kalhauge2019binary,niemetz2013ddsmt}
have been proposed to automate test input minimization.
Delta debugging algorithms, \eg, \ddmin and \probdd,
are among the most general and widely used techniques, upon which many advanced tools
such as  \hdd and \perses are built.
Since our approaches, \ie, \wddmin and \wprobdd, are the improved
versions of \ddmin and \probdd, respectively, we first explain
the workflows of \ddmin and \probdd with an example.

\cref{subfig:example} displays a program that triggers a
real-world compiler bug GCC-71626~\cite{gcc71626report}.
It triggers GCC to crash when compiling the program.
We aim to minimize this program to the smallest size
while still triggering the compiler bug, thus facilitating debugging.
Taking the program as plain text and performing delta debugging algorithms
on it directly is inefficient, as the program is highly structured.
In practice, delta debugging is usually wrapped in tree-based
techniques, \eg, \hdd and \perses, being applied on the tree level.
In the tree representation of the program, \eg, the parse tree,
there are eight nodes at the same level right under the root node (highlighted in orange in \cref{subfig:example}),
each corresponding to a distinct part of the program such as a
\mycode{typedef} statement, or a function definition.
To minimize the program, both \hdd and \perses
invoke \ddmin or \probdd to minimize the tree nodes
starting from this level, \ie, $[1,2,3,4,5,6,7,8]$.

\subsection{Workflow of \ddmin}
\label{subsubsec:ddmin}

Given $\testInput$ and  \property, \ddmin works in the following steps.

\noindent\underline{Step 1:}
Split $\testInput$ into $n$ partitions evenly by length.
For each partition $\p$, test if $p$ alone preserves \property, \ie, $\property(\p)=\AlgTrueLiteral$.
If yes, remove all other partitions from $\testInput$ and resume Step 1 with $n=2$; otherwise, go to Step 2.

\noindent\underline{Step 2:}
Test if the complement of each partition $\p$ preserves \property, \ie, $\property(\testInput\setminus\p)=\AlgTrueLiteral$.
If yes, remove $\p$ from $\testInput$ and resume Step 1 with $n=n-1$; otherwise, go to Step 3.

\noindent\underline{Step 3:}
Terminate if each partition $\p$ contains only one element; otherwise, double $n$ and resume Step 1.

Starting from $n=2$ and following the above steps,
\ddmin performs \ddminQueriesOnExample property tests
in total to minimize the program in \cref{subfig:example}.
The specific property tests \ddmin performs during the minimization process
are shown in \cref{subfig:ddmin-iterations}.
Note that
\ddmin may produce duplicate test inputs, which are not listed in the
figure, since in practice they can be recognized and
skipped by caching the tests that have been performed~\cite{zeller2002simplifying,rcc}.

\begin{figure*}[h!]
    \centering
    \begin{subfigure}[b]{0.48\columnwidth}
        \includegraphics[width=\columnwidth]{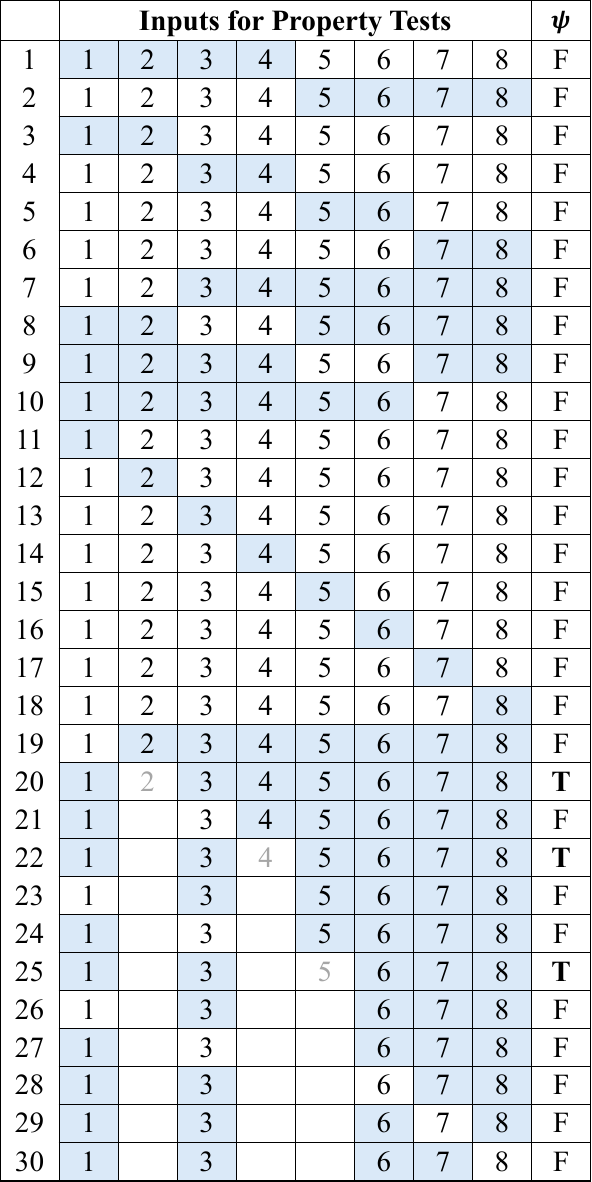}
       \caption{
       	\ddmin
       	}
       \label{subfig:ddmin-iterations}
    \end{subfigure}
        \hfil
    \begin{subfigure}[b]{0.48\columnwidth}
        \includegraphics[width=\columnwidth]{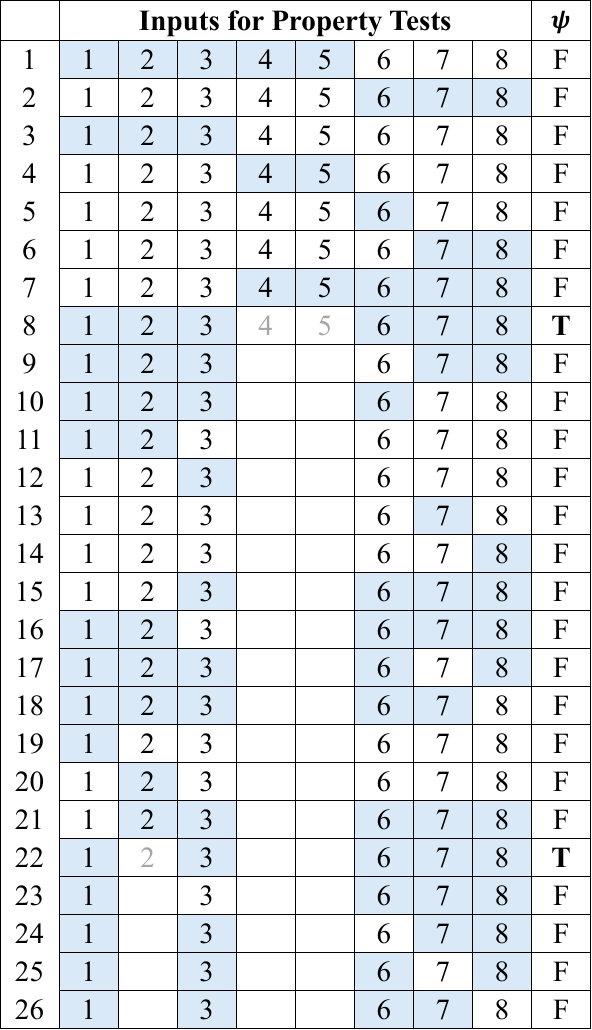}
        \caption{\wddmin
        	}
        \label{subfig:wdd-iterations}
    \end{subfigure}
        \hfil
    \begin{subfigure}[b]{0.48\columnwidth}
        \includegraphics[width=\columnwidth]{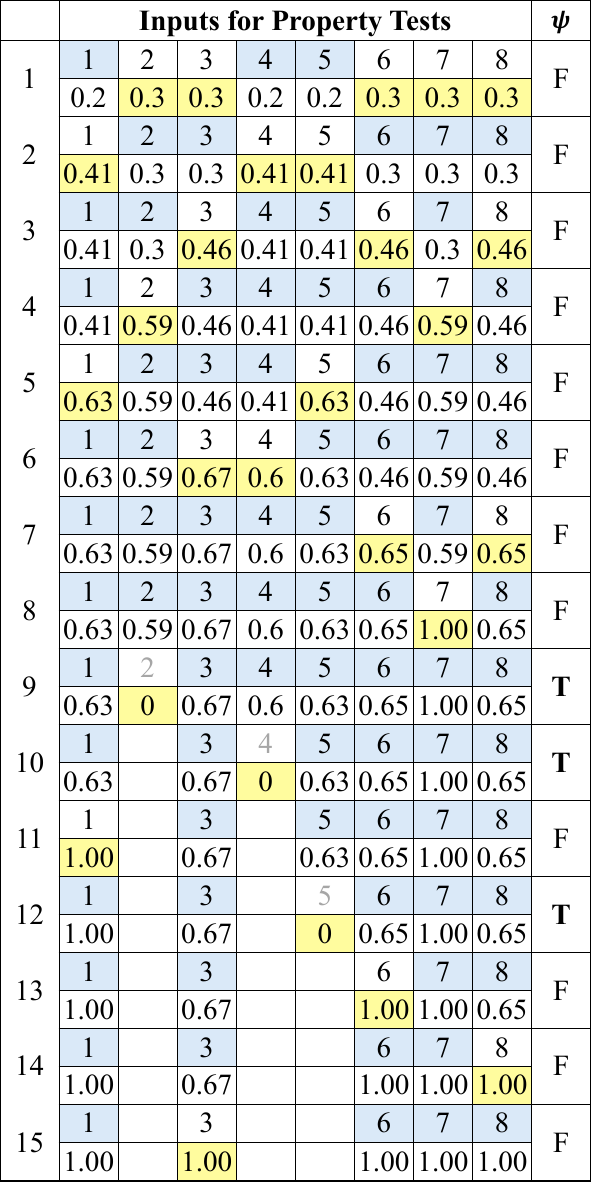}
        \caption{
            \probdd
        }
        \label{subfig:probdd-iterations}
    \end{subfigure}
        \hfil
    \begin{subfigure}[b]{0.48\columnwidth}
        \includegraphics[width=\columnwidth]{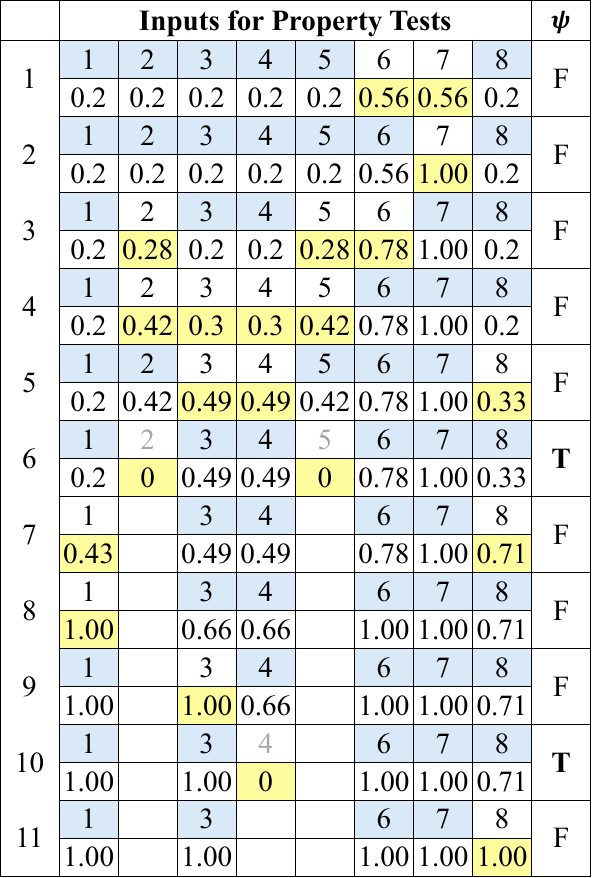}
        \caption{\wprobdd
        	}
        \label{subfig:wprobdd-iterations}
    \end{subfigure}
    \caption{The detailed minimization process of \ddmin, \wddmin, \probdd, and \wprobdd.
        The elements selected for the property test in each iteration are highlighted in blue,
        with the leftmost column indicating the index of each property test.
        In \cref{subfig:probdd-iterations} and \cref{subfig:wprobdd-iterations},
        the probabilities updated after each test are highlighted in yellow.
        The last column of each figure shows
        the result of the property test \property. In this case, all
        the four algorithms minimize the input list to the same result, which is
        $[1,3,6,7,8]$.
    }
    \label{fig:dd-iterations}
\end{figure*}

\subsection{Workflow of \probdd}
\label{subsec:probdd}
Different from \ddmin, which follows a predefined pattern
to perform the deletion operations, \probdd~\cite{wang2021probabilistic}
employs a probabilistic model to guide the entire minimization process.
The key insight of \probdd is to estimate
the probability of each element appearing in the minimized
result with a probabilistic model.
Given $\testInput$ and \property,
and a map $\probs$ that stores the estimated probabilities of
each element in $\testInput$ appearing in the minimized result
(the initial probability of each element is set to a same value, \eg, 0.2),
\probdd works in the following steps.

\noindent\underline{Step 1:}
Sort the elements in $\testInput$ in ascending order of their probabilities.
Select a prefix $\prefix$ from the sorted list that maximizes
the expectation of the number of elements that can be removed,
\ie, $|\prefix| \times \prod_{\e \in \prefix}^{} (1-\probs[\e])$.

\noindent\underline{Step 2:}
Test if the complement of $\prefix$ preserves \property,
\ie, $\property(\testInput\setminus\prefix)=\AlgTrueLiteral$.
If yes, remove $\prefix$ from $\testInput$, set the probabilities of the elements in $\prefix$ to 0,
and go to Step 4;
if not, go to Step 3.

\noindent\underline{Step 3:}
Increase the probabilities of elements in $\prefix$ according
to the probabilistic model~\cite{wang2021probabilistic}, then go to Step 4.

\noindent\underline{Step 4:}
Terminate if the probabilities of all the elements in $\testInput$
reach 1; otherwise, go to Step 1.

Following the above steps,
the minimization process of the example program in \cref{subfig:example}
is shown in \cref{subfig:probdd-iterations}.
Each property test is represented with two rows, where the first row
displays the elements selected (complement of $\prefix$) for testing,
and the second row shows the probability of each element after the test.
The selected elements and the updated probabilities are
highlighted with blue and yellow, respectively. Starting from
the same initial probability (set to 0.2 in this case),
\probdd performs \probddQueriesOnExample property
tests to finish the minimization process.

\subsection{1-Minimality}
\label{subsec:1-minimality}
The ultimate goal of test input minimization is to obtain the
globally minimal result, where no smaller input can exhibit \property.
However, previous work has proven that obtaining the global
minimality is NP-complete~\cite{zeller2002simplifying,misherghi2006hdd}.
In practice, the goal is usually
relaxed to local minima. First presented by
\dd~\cite{zeller2002simplifying}, 1-minimality has been widely
adopted by a series of works~\cite{xu2023pushing,misherghi2006hdd,sun2018perses,heo2018effective}
as the criterion of minimality evaluation.
A minimized input is considered 1-minimal if no single element
can be further removed without losing the property \property.
\hdd~\cite{misherghi2006hdd} extends the principle of 1-minimality
to tree structures, introducing 1-tree-minimality, which promising
that, in the tree representation of the input, no single tree node
can be further removed without violating the property.
To achieve 1-tree-minimality, tree-based techniques, \eg, \hdd~\cite{misherghi2006hdd}
and \perses\cite{sun2018perses}, typically operate in a \emph{fixpoint mode}.
In this mode, the minimization process is repeatedly applied to the minimized result
until no more tree nodes can be removed from the result.

\section{Motivation}
\label{sec:motivation}
As \cref{subfig:example} shows,
the code snippets represented by different nodes vary in size.
For example, while node \circled{1} represents a \mycode{typedef}
statement containing \nodeOneTokens tokens, node \circled{7} defines the
function \mycode{test2llong1} with \nodeSevenTokens tokens.
This discrepancy in size of nodes can affect the efficiency and
effectiveness of minimization. However, both \ddmin and \probdd
fail to capture this information and treat all nodes uniformly,
thus leaving room for improvement. This is where out concept of
\wdd comes into play. The key insight of \wdd is to assign each
element a weight that matches its size, and perform weight-based partitioning.
We first define the weight
of elements in delta debugging, based on which, we present two new
delta debugging algorithms, \wddmin and \wprobdd, by applying
\wdd to \ddmin and \probdd, respectively.

\begin{definition}[Weight] \label{def:weight}
	The weight of an element in the input list of delta debugging is
    defined as the size of the fragment represented by the element.
    The weight of a partition is the sum of the weights of all elements
    in the partition. The size is typically measured by the number of tokens.
\end{definition}

\subsection{Improving \ddmin}
\cref{subfig:dd-searchspace} visualizes the search
space of \ddmin in a tree, illustrating that \ddmin splits
the list evenly by length to conduct a binary search-style
deletion. However, it fails to achieve the true evenness
due to the effect of different weights of nodes.
As highlighted in orange in \cref{subfig:dd-searchspace}, the weights
of partitions on each level vary significantly, which
can impair the efficiency of \ddmin. That is because,
statistically speaking, a larger partition is more likely
to contain the failure-inducing elements, and thus less likely
to be removed. However, \ddmin fails to capture this information
and handles all nodes equally, leading to its efficiency being
hampered by spending a large amount of attempts on deleting nodes
that are unlikely to be successfully removed. For instance, the largest
node (node \circled{7}) in the previous example, which is the core
element to trigger the compiler bug, is attempted to be removed
from the list with partitions for 13 times during the minimization.

Different from \ddmin, \wddmin considers the weights of elements and
performs a weight-based partitioning to
make the actual size of each partition as close as possible.
The search space of \wddmin based on this strategy is shown
in \cref{subfig:wdd-searchspace}. Following this search space, \wddmin
finish the minimization of the example program with only
\wddQueriesOnExample property tests, and attempts to remove node \circled{7}
only 12 times. The detailed minimization process is shown in \cref{subfig:wdd-iterations}.
This improvement is much more significant for larger
and more complex inputs, as demonstrated in \cref{subsubsec:rq2-efficiency}.

\subsection{Improving \probdd}
As described in \cref{subsec:probdd}, \probdd strives to
maximize the expectation of the number of elements
that can be removed during partitioning. However, the number of
elements does not necessarily correspond to the number of tokens
that can be deleted. For example, given two elements with the same
probability of being removed, the one with more tokens (\ie, larger weight)
should be chosen to remove first, since deleting it contributes more to
global minimization process. The performance of \probdd is
suboptimal since it fails to consider the weight of elements
when constructing the probabilistic model.
To fill this gap, \wprobdd leverages the weight information of elements
to refine the probabilistic model of \probdd, and uses this model to guide
partitioning.
As shown in \cref{subfig:wprobdd-iterations},
boosted by the weighted model, \wprobdd minimizes the example program
with only \wprobddQueriesOnExample property tests.
It is worth clarifying that, although in this example,
\probdd and \wprobdd produce the same minimized result,
our evaluation in \cref{sec:evaluation} demonstrates the
superior effectiveness of \wprobdd over \probdd in practice
by producing smaller minimized results.

\section{Weighted Minimizing Delta Debugging}
\label{sec:approach-wdd}
This section describes the application of \wdd
to improve the efficiency of \ddmin.
\cref{alg:weightdd} details \wddmin,
with our extensions beyond \ddmin highlighted with
grey blocks.
Compared to \ddmin, \wddmin has a different
partitioning strategy \FuncWeightPartition on \cref{alg:weightdd:weightedPartition-begin},
and an additional deletion pass \FuncCheckOneMinimal on \cref{alg:weightdd:checkOneMinimal:start}
to ensure 1-minimality.

Started with the whole input \testInput{} as the only partition (\cref{line:wdd:init:partition}),
\wddmin performs systematic deletion operations on
the partitions and their complements, and iteratively
splits the partitions into smaller ones.
If a partition \partition alone preserves the property
(\ie, $\property(\partition)$ on \cref{line:wdd:propertytest:partition}),
all the other partitions
are removed, and the algorithm restarts with this single remaining partition
(\cref{alg:weightdd:testPartitionStart}-\ref{alg:weightdd:testPartitionEnd}).
If the complement of a partition exhibits the property (\ie, $\property(\complement)$
on \cref{line:wdd:propertytest:complement}),
\wddmin removes the partition and restarts with the remaining partitions
(\cref{alg:weightdd:testComplementStart}-\ref{alg:weightdd:testComplementEnd}).
If no partition or complement exhibits \property,
\wddmin calls
\FuncWeightPartition (\cref{alg:weightdd:increase-granularity}) to split
the partitions into smaller ones
based on the weights of the elements in these partitions,
and then start a new iteration.
This process terminates when the partition list \partitions is empty (\cref{line:wdd:bigloop:condition}).
Then \wddmin performs an additional deletion pass by calling
\FuncCheckOneMinimal (\cref{alg:weightdd:checkOneMinimal}) to
make sure the produced result is 1-minimal.
\begin{algorithm}[h!]
    \footnotesize
    \DontPrintSemicolon
    \SetKwInput{KwData}{Input}
    \SetKwInput{KwResult}{Output}
    \caption{
        \weightedMinimizingDeltaDebugging
    }
    \label{alg:weightdd}

    \KwData{$\testInput\in\searchspace$: the input list of elements.}
    \KwData{\HiLi{-85pt}{$\weight:\elementspace\rightarrow N$: the weights of each element.}}
    \KwData{$\property: \searchspace \rightarrow \boolspace$: the property to be preserved. }
    \KwResult{the minimized list that preserves the property.}

    \BlankLine

    $\testInputMin \gets \testInput$ \;
    $\partitions \gets [\testInput]$ \label{line:wdd:init:partition} \;
    \testInputMin $\gets$ \FuncWDD{$\partitions, \testInputMin, \property, \weight$}\;
    \KwRet \HiLi{-126pt}{\FuncCheckOneMinimal{$\testInputMin, \property$}\;} \label{alg:weightdd:checkOneMinimal}

    \BlankLine

    \SetKwProg{Fn}{Function}{:}{}
    \Fn{\FuncWDD{\partitions, \testInputMin, \property, \weight}}{
        \While{$|\partitions|\not =0$}{ \label{line:wdd:bigloop:condition}
            \ForEach{$\partition \in \partitions$}{ \label{alg:weightdd:testPartitionStart}
                \If{$\property \xspace (\partition)$}{\label{line:wdd:propertytest:partition}
                    \testInputMin $\gets$ \partition\;
                    \partitions $\gets$ \FuncWeightPartition{$[\partition]$, \weight}\;
                    \KwRet{\FuncWDD{\partitions, \testInputMin, \property, \weight}}\;
                }
            } \label{alg:weightdd:testPartitionEnd}
            \ForEach{$\partition \in \partitions$}{ \label{alg:weightdd:testComplementStart}
                $\complement \gets \testInputMin \setminus \partition     $\;
                \If{$\property \xspace (\complement)$}{\label{line:wdd:propertytest:complement}
					$ \testInputMin \gets \complement $ \;
					$ \partitions \gets \partitions \setminus [\partition]  $ \;
                    \KwRet{\FuncWDD{\partitions, \testInputMin, \property, \weight}}\;
                }
            } \label{alg:weightdd:testComplementEnd}
            \partitions $\gets$ \HiLi{-79pt}{\FuncWeightPartition{\partitions, \weight}\;} \label{alg:weightdd:increase-granularity}
        }
        \KwRet{\testInputMin}\;
    }

    \Fn{ \HiLi{-107pt}{\FuncWeightPartition{\partitions, \weight}} }{ \label{alg:weightdd:weightedPartition-begin}
        $\newPartitions \gets       \emptylist     $ \;
        \ForEach{$\partition \in \partitions$}{
            \lIf{$|\partition| = 1$}{ \label{alg:weight-partition:single-node-begin}
                \Continue
               \tcp*[h]{skip this partition} \label{alg:weight-partition:single-node-end}
            }

            \halfWeight $\gets$ $0.5 \times \sum_{e \in \partition}^{} \weight(e)$  \;  \label{alg:weight-partition:split-begin}
            \pOne, \pTwo $\gets$ split \partition into two partitions    with weight sum of each close to \halfWeight\;
            $ \newPartitions \gets \newPartitions + [\pOne, \pTwo ] $
\tcp*[h]{add \pOne, \pTwo to \newPartitions} \label{alg:weight-partition:split-end}

        }
        \KwRet{\newPartitions}\; \label{alg:weightdd:weightedPartition-end}
    }

    \Fn{ \HiLi{-122pt}{\FuncCheckOneMinimal{\testInputMin, \property}} }{ \label{alg:weightdd:checkOneMinimal:start}
        loopStart:
        \ForEach{$\node \in \testInputMin$}{   \label{alg:weightdd:checkOneMinimal:loopStart}
            $\complement \gets \testInputMin \setminus [\node]$\;
            \If{$\property(\complement)$}{
                $ \testInputMin \gets \complement  $\;
                goto loopStart\;
            }
        } \label{alg:weightdd:checkOneMinimal:loopEnd}
        \KwRet{\testInputMin}\; \label{alg:weightdd:checkOneMinimal:end}
    }

\end{algorithm}

\subsection{Weighted Partitioning Strategy}
The main extension of \wddmin is the partitioning strategy, as shown
in function \FuncWeightPartition (\cref{alg:weightdd:weightedPartition-begin}-\ref{alg:weightdd:weightedPartition-end}).
Unlike \ddmin, which partitions the input list \testInput
evenly by the \emph{number} of elements,
\wddmin aims to split \testInput evenly by the \emph{weight} of elements,
striving to \emph{make the weight of each partition as close as possible}.
(\cref{alg:weight-partition:split-begin}-\ref{alg:weight-partition:split-end}).
Notably, if a partition from the current iteration contains only
one element, the partition will be excluded from the partition list in the next iteration (\cref{alg:weight-partition:single-node-begin}),
because the partition cannot be further divided.

Revisiting the example in \cref{subfig:example}, by applying
the weight-based partitioning strategy, the search space is reorganized
as shown in \cref{subfig:wdd-searchspace}. While the tree is not
balanced in terms of the number of elements, it achieves balance
for the weight of each partition.
Quantitatively, \wddmin
strives to minimize the standard deviation of the
weights of partitions during partitioning. For example, in the second iteration
(corresponding to the third level of the tree in \cref{subfig:dd-searchspace} and \cref{subfig:wdd-searchspace}),
the standard deviation of the partition weights of \ddmin (\ie, $[13, 14, 24, 31]$) is \ddminExampleStdDev,
whereas that of \wddmin (\ie, $[20, 15, 16, 31]$) is only \wddExampleStdDev.

\subsection{1-Minimality of \wddmin}
\label{subsec:1-minimality-of-wdd}
\wddmin guarantees 1-minimality with an additional deletion pass,
as shown in function \FuncCheckOneMinimal
(\cref{alg:weightdd:checkOneMinimal:start}-\ref{alg:weightdd:checkOneMinimal:end}).
Because of the weight-based partitioning strategy, larger elements are isolated earlier in the deletion process.
For example, in \cref{subfig:wdd-searchspace},
node \circled{6} is isolated as a separate partition in the third iteration,
and it cannot be removed in the current iteration. However, in practice, the
deletion of some nodes may benefit the deletion of other
nodes~\cite{zeller2002simplifying,misherghi2006hdd,regehr2012test}.
To ensure 1-minimality,
\wddmin attempts to remove each
remaining element individually in the end by calling function
\FuncCheckOneMinimal (\cref{alg:weightdd:checkOneMinimal}).
The loop (\cref{alg:weightdd:checkOneMinimal:loopStart}-\ref{alg:weightdd:checkOneMinimal:loopEnd})
iteratively checks whether each remaining element can be removed
without losing the property.
If so, the element
is removed and the loop restarts. This process
continues until no element can be further removed,
so that 1-minimality is guaranteed.

\subsection{Time Complexity of \wddmin}
\wddmin does not shrink or enlarge the search space of \ddmin.
Instead, \wddmin
follows the similar deletion process as \ddmin with a more rational partitioning strategy.
Therefore, by design, \wddmin has the same worst-case time complexity
as \ddmin, \ie, $O(n^2)$~\cite{zeller2002simplifying},
where $n$ is the number of elements in the input list.

\myparagraph{Average Time Complexity}
We argue that \wddmin can achieve higher overall efficiency
than \ddmin in practice.
The key insight of
\wddmin is that the probability of an element being removed
varies with its weight,
and there is a negative correlation between them.
Intuitively, an element with a larger weight, \ie, representing a larger
fragment of a test input, is less likely to be removed than a smaller one,
as it is more likely to contain the failure-inducing elements.
The statistical validation of this observation is provided in \cref{subsec:rq1}.
With this insight,
we expect that \wddmin can achieve better efficiency than \ddmin.
We perform a simulation below to demonstrate this.

\subsection{Synthetic Analysis for Average Time Complexity}
The inherent complexity of delta debugging problem prevents us
from proving \wddmin is better than \ddmin in all cases, which is
also not necessarily true in practice. Therefore, we design this simulation to compare the efficiency of \wddmin and \ddmin.

\subsubsection{Analysis Setup}
First, we randomly synthesize a set of lists and predetermine their minimization results.
Next, we perform \wddmin and \ddmin on the lists and record the numbers of property tests
required by each algorithm on each list respectively.
The minimization results are predetermined based on the probability of each element being removed,
and the probabilities are calculated based on the assumption below.

\begin{Assumption}[Randomness] \label{Assumption:randomness}
	For a random input, each token has the same probability of being removed.
\end{Assumption}

Given this assumption and the probability of a token being removed $p_0$,
the probability of an element with $w$ tokens being removed $p_e$ equals to $p_0^{w}$.
That is because an element can be removed only if all its tokens can be removed.
With the input lists of elements synthesized randomly, this assumption
helps quantitatively distinguish the probabilities of elements with different weights being removed,
so that we can predetermine the minimization result.
This assumption is not necessary for the correctness of \wddmin in practice.

With the above assumption, we perform the simulation as follows.
To synthesize a random input list, we first generate a length $n$ of the list,
where $n$ is a random integer between 2 and 1,000 (\ie, $n \in [2, 1000]$),
and the total number of tokens represented by the elements in the list,
which is a random integer between $n$ and $10n$.
The number of tokens for each element is distributed randomly, for instance,
a list of length 4 with 10 tokens could be \mycode{[1, 3, 2, 4]}. To predetermine
the minimization result, we first generate a random value $p_0 \in (0, 1)$, which
represents the probability of each token being removed. Then, we calculate the
probability of each element being removed $p_e$ based on \cref{Assumption:randomness}.
After that, we generate a random value $p \in [0, 1]$ for each element,
and compare it with $p_e$ to determine whether the element can be removed.
The element can be removed if $p < p_e$, otherwise, it cannot be removed.
Based on the established result, the property \property is preserved if all the non-removable
elements are included in the list.
We execute \wddmin and \ddmin to minimize the synthesized list,
and record their numbers of property tests during the minimization process, respectively.
The effect of randomness is eliminated by
repeating the single process for a large number of times.
Specifically, we perform \ddmin and \wddmin on 5,000 randomly synthesized lists.

\begin{figure}[h!]
    \centering
    \includegraphics[width=0.8\columnwidth]{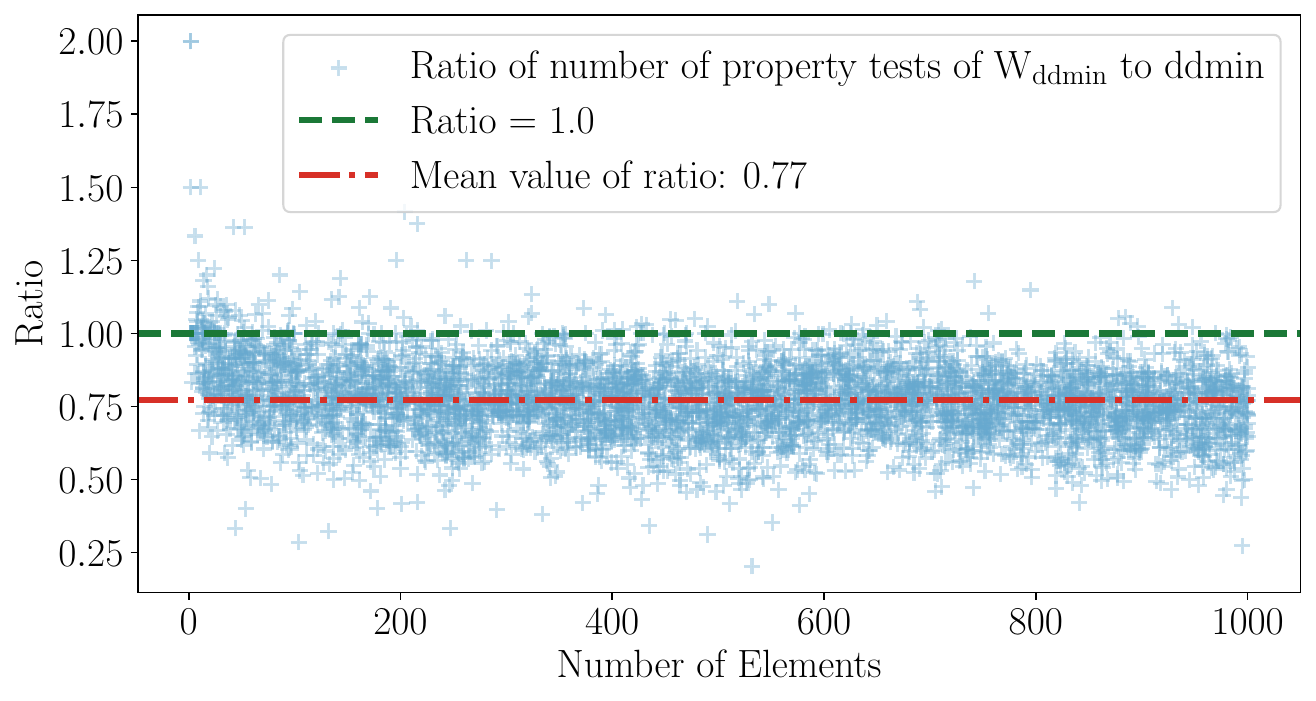}
    \caption{
        The simulation results of \wddmin and \ddmin on the synthetic data.
        }
    \label{fig:wdd-simulation}
\end{figure}

\subsubsection{Analysis Result}
The detailed results are shown in \cref{fig:wdd-simulation}.
On average, \wddmin uses 23\% fewer property tests than \ddmin
to finish the minimization. The results emulatively demonstrate
the superior efficiency of \wddmin compared to \ddmin in the ideal
case where the probabilities of elements being removed are
negatively correlated with their weights.
We verify this correlation and evaluate the practical efficiency
of \wddmin on real-world benchmarks in \cref{subsec:rq2}.

\section{Weighted Probabilistic Delta Debugging}
\label{sec:approach-wprobdd}
\begin{algorithm}[h]
    \footnotesize
    \DontPrintSemicolon
    \SetKwInput{KwData}{Input}
    \SetKwInput{KwResult}{Output}
    \caption{
    	\weightedprobabilisticdeltadebugging
    	}
    \label{alg:weightprobdd}
    \newcommand{\boxInitProbs}[1]{\tikz[remember picture,overlay]{\node[yshift=3pt,xshift=5pt,fill=#1,opacity=.25,fit={(A)($(B)+(.9\linewidth,\baselineskip)$)}] {};}\ignorespaces}
    \newcommand{\boxGetNextTest}[1]{\tikz[remember picture,overlay]{\node[yshift=3pt,xshift=5pt,fill=#1,opacity=.25,fit={(A)($(B)+(.9\linewidth,.3\baselineskip)$)}] {};}\ignorespaces}
    \KwData{$\testInput\in\searchspace$: the input list of elements.}
    \KwData{\HiLi{-85pt}{$\weight:\elementspace\rightarrow N$: the weights of each element.}}
    \KwData{$\property: \searchspace \rightarrow \boolspace$: the property to be preserved.}
    \KwData{$\pinit$: the initial probability for each element.}
    \KwResult{the minimized list that preserves  \property.}
    \BlankLine

    $\testInputMin \gets \testInput$ \;
    $\probs \gets \{ n \rightarrow \pinit | n \in \testInput  \}$ \tcp{the probability function that records and returns the probability of each element in \testInput}
    \While{not \FuncShouldTerminate{\probs} }{
        \partition $\gets$ \HiLi{-73pt}{\FuncGetNodesToRemove{\testInputMin, \probs, \weight}\;}
        \complement $\gets$ \testInputMin $\setminus$ \partition\;
        \lIf{$\property \xspace (\complement)$}{
        	$\testInputMin \gets \complement$
        }
        \lElse{
            \probs $\gets$ \FuncUpdateProbability{\partition, \probs}
        }
    }
    \KwRet{\testInputMin}\;

    \BlankLine

    \Fn{\HiLi{-86pt}{\FuncGetNodesToRemove{\testInputMin, \probs, \weight}} }{ \label{alg:weightprobdd:getNextTest:begin}
        $\sortedTestInput \gets$ sort the elements in \testInputMin by the value of $\weight(\node) * (1 - \probs(\node))$ in descending order\; \label{alg:weightprobdd:sort}
        $\newPartitions \gets \emptylist$, $\partition \gets \emptylist$, $\maxGain \gets 0$\; \label{alg:weightprobdd:getNextTest:begin2}
        \ForEach{\node\ $\in$ \sortedTestInput}{
        	$\partition \gets \partition + [\node]$\;
            \weightFullTerm $\gets$ $\sum_{n_i \in \partition}^{} \weight(n_i)$\;
            \probOfDeletion $\gets$ $\prod_{n_j \in \partition}^{} (1-\probs(n_j))$\;
            \gain $\gets$ \weightFullTerm $\times$ \probOfDeletion\;
            \If{\gain $>$ \maxGain}{
                \maxGain $\gets$ \gain\;
                \newPartitions $\gets$ \partition\;
            }
        }
        \KwRet{\newPartitions}\; \label{alg:weightprobdd:getNextTest:end}
    }

    \Fn{\FuncShouldTerminate{\probs}}{
        \tcp*[h]{Implementation skipped. Same as \probdd in \cite{wang2021probabilistic}.}
    }

    \Fn{\FuncUpdateProbability{\partition, \probs}}{
        \tcp*[h]{Implementation skipped. Same as \probdd in \cite{wang2021probabilistic}.}
    }

\end{algorithm}

To demonstrate the generality of \wdd, we applied
the concept of \wdd to improve \probdd (a representative
variant of \ddmin) and thus proposed a new minimization algorithm \wprobdd.
As described in \cref{subsec:probdd},
with the model that tracks the expected probability of each element remaining in the result,
the partitioning principle of \probdd is to prioritize the deletion of elements
with lowest probability and maximize the expected number of elements that
can be successfully removed.
However, the ultimate goal of the minimization is to delete the most tokens
possible, instead of the most elements. Due to the different
sizes of elements, there is a gap between the principle of \probdd
and the ultimate goal of the minimization, which makes \probdd suboptimal.
To bridge this gap,
\wprobdd improves \probdd by
incorporating the weight of elements as a new factor into the probabilistic model.

\cref{alg:weightprobdd} shows the workflow of \wprobdd,
and
the key extensions beyond \probdd are highlighted with grey blocks.
When deciding the partition to remove in each test (implemented in function \FuncGetNodesToRemove),
the fundamental principle of \wprobdd is to
\emph{(1) prioritize the deletion of elements that are likely to remove larger weight}, and
\emph{(2) maximize the expected value of weight that can be removed.}
To realize the principle, \wprobdd first sorts the elements in the list in descending order,
by the expectation of the value of weight that can be removed by
attempting to delete the element.
This value equals to the product of the probability
of the element can be removed and the value of its weight (\cref{alg:weightprobdd:sort}).

After that, \wprobdd determines the partition to remove in each test with the sorted list.
Technically, starting from the first element in the sorted list,
\wprobdd can include any number of elements in the partition to remove in the next test.
While including more elements increases value of weight that can be removed,
it also decreases the probability of the test passing.
To balance the trade-off, \wprobdd chooses the number of elements
for removal that maximizes the expectation of the value of weight
that can be removed successfully. To this end, \wprobdd redefines
the gain function in \probdd with the weights of elements as
$
Gain(m) = \sum_{i=1}^{m} w_i \cdot \prod_{j=1}^{m} (1 - p_{j})
$
where $m$ is the number of elements to be removed,
$w_i$ is the weight of the $i$-th selected element,
and $p_j$ is the probability of being remained of the $j$-th element.
As shown in function \FuncGetNodesToRemove
(\cref{alg:weightprobdd:getNextTest:begin2}-\ref{alg:weightprobdd:getNextTest:end}),
\wprobdd selects a certain prefix of the sorted list \sortedTestInput that maximizes the
gain function as the partition, and attempts to remove this partition
in the next test. The complexity of this process is $O(n)$, where
$n$ is the length of the list.

The rest steps of \wprobdd are similar to \probdd, including
performing property tests, and updating the probabilities of
elements according to prior test results.
We exclude the explanation of these steps here,
instead, and refer the readers to the original paper of
\probdd~\cite{wang2021probabilistic} for details.

\subsection{Minimality of \wprobdd}
\wprobdd promises the same minimality as \probdd, which is conditional 1-minimality.
The result of \probdd is 1-minimal under the assumption that
the deletability of each element is independent. However, this
assumption typically does not hold in practice, since the deletion of some
elements may affect the deletability of other elements. For example,
even if a statement that defines a variable is bug-irrelevant,
it can only be removed after all the statements that use the variable
are removed.
Despite sharing the same minimality,
\wprobdd is expected to generate smaller results than \probdd,
since \wprobdd always strives to maximize the weight (\ie, the number of tokens)
that can be removed in the next test.
We evaluate the practical effectiveness of \wprobdd in \cref{subsec:rq2}.

\subsection{Time Complexity of \wprobdd}
\wprobdd shares the same worst-case time complexity as \probdd,
which is $O(n)$~\cite{wang2021probabilistic}, where $n$ is
the length of the input list. In practice, the deletion strategy of \wprobdd
\ie, maximizing the expected weight can be removed, not only
enhances effectiveness, but also speeds up the minimization process.
That is because, successfully removing a partition containing
a large number of tokens can usually make the execution of subsequent
tests faster. Therefore, we expect that \wprobdd can
outperform \probdd in terms of time efficiency.
This expectation can hardly be verified by a simulation,
so we directly evaluate the efficiency of \wprobdd on real benchmarks in \cref{subsec:rq2}.

\section{Evaluation}
\label{sec:evaluation}

In this section, we verify the significance of \wdd by
evaluating the effectiveness and efficiency of \wddmin and \wprobdd.
we explore to what extent
\wddmin and \wprobdd outperform \ddmin and \probdd
in different application scenarios, respectively.
We select \hdd and \perses
for evaluation as they are two state-of-the-art test input
minimization tools that rely on delta debugging.
For each of the two techniques, we
implement \wddmin and \wprobdd versions to replace their original
versions with \ddmin and \probdd, respectively, and compare their
performance with the original versions.
For ease of presentation, we refer to \hdd with \ddmin, \wddmin,
\probdd and \wprobdd as \hddDD, \hddWDD, \hddProbDD, and \hddWProbDD,
respectively. Similarly, the four versions of \perses are referred
to as \persesDD, \persesWDD, \persesProbDD, and \persesWProbDD, respectively.
All the minimization techniques for evaluation are executed in the fixpoint mode
as described in \cref{subsec:1-minimality}.
For fair comparison,
all experiments were conducted on an Ubuntu 22.04 server with an Intel
Xeon
CPU @ 2.60GHz and 512 GB RAM,
using a single-process,
single-threaded.

We aim to answer the following research questions.
\begin{enumerate}[topsep=0pt,leftmargin=*]
    \item  What is the correlation between element weight and the probability of being removed in practice?
    \item  How does the performance of \wddmin compare to \ddmin?
    \item  How does the performance of \wprobdd compare to \probdd?
\end{enumerate}

\myparagraph{Benchmarks}
We conducted experiments with \allBenchmarksSize benchmarks.
Each benchmark triggers a real-world bug in a certain language processor,
and is considerably large and complex,
aligning with real-world application scenarios of test input minimization.
Specifically, we utilized the following benchmarks.
\begin{itemize}[leftmargin=*,topsep=0pt]
    \item \textbf{C}: We collected \cBenchmarkSize C programs
    from previous studies~\cite{sun2018perses,zhang2023ppr,rcc}.
    These programs trigger real bugs in \llvm and \gcc,
    and are large, complex with \cProgramsAverageTokens tokens on average.

    \item \textbf{XML}:
    To increase the diversity of the benchmark suite,
    we included \xmlBenchmarkSize XML files, with each triggering
    a bug in Basex~\cite{basex},
    a widely used XML database and Xquery processor.
    These benchmarks are also large and complex,
    containing \xmlProgramsAverageTokens tokens on average.
\end{itemize}

\myparagraph{Metrics}
\label{subsec:metrics}
We used the following metrics to evaluate different algorithms,
following \cite{sun2018perses,xu2023pushing,rcc,zhang2023ppr,wang2021probabilistic}.
\begin{itemize}[leftmargin=*,topsep=0pt]
    \item \textbf{S(\#)}: the number of tokens in the minimized result.
        A lower value means a more effective minimization by removing more property-irrelevant elements.
    \item \textbf{T(s)}: the processing time in seconds.
        Shorter time means higher efficiency.
    \item \textbf{Speed}: the number of tokens deleted per second.
        Using processing time to gauge efficiency is not comprehensive,
        for cases where one approach generates a smaller result but also takes longer time.
        We measure the number of tokens
        deleted per second to balance the trade-off
        between effectiveness and time consumption.
    \item \textbf{Wilcoxon signed-rank test}~\cite{woolson2005wilcoxon}:
        to measure the statistical significance of the improvements our our approaches.
        A small p-value (typically $<0.05$) from this test
        suggests a statistically significant difference between the paired data.

\end{itemize}

\subsection{RQ1:  Element Weight \vs Deletion Probability Correlation}
\label{subsec:rq1}
The first question we are curious about is the correlation between the weights of elements
and their probabilities of being removed.
Since the fundamental observation behind \wddmin is that larger elements
are less likely to be deleted than smaller ones,
we would like to verify if our assumption, \ie,
\textit{the probability of elements being deleted is negatively correlated with their weights},
holds during the execution of \ddmin in practice. Specifically, for an input list \testInput and its minimized
result \testInputMin, the probability of elements with weight \weight
being deleted $\pDel(\weight)$ is defined as the ratio of the number of elements with weight \weight
that are deleted to the total number of elements with weight \weight, \ie,
$
\pDel(\weight) = \frac{\#(\weight, \testInput) - \#(\weight, \testInputMin)}{\#(\weight, \testInput)}
$,
where $\#(\weight, \testInput)$ denotes the number of elements with weight \weight in list \testInput.
To evaluate the correlation,
we calculate the Spearman's rank correlation coefficient~\cite{zar2005spearman}
between the probabilities of elements being deleted and their weights
for each execution of \ddmin.
Being widely used in practice~\cite{chok2010pearson,ali2022spearman},
Spearman's rank correlation coefficient $\rho$~\cite{zar2005spearman}
is a non-parametric measure of the strength and direction of association between two ranked variables.
The value of $\rho$ ranges from -1 to 1:
$\rho = 1$ indicates a perfect positive correlation,
$\rho = -1$ indicates a perfect negative correlation,
and $\rho = 0$ implies no correlation.

To answer this research question, we use \hddDD and \persesDD to minimize the test inputs in our benchmarks,
and record the weights of elements before and after each execution of \ddmin.
Cases where no elements are removed are excluded,
since $\rho$ is undefined in these scenarios.
We then calculate $\rho$ for each execution of \ddmin.
Since \ddmin is normally performed multiple times when minimizing a test input,
the $\rho$ of each benchmark is calculated as the average of the $\rho$ values from all executions
of \ddmin for that benchmark.

\begin{figure}[ht]
    \centering
    \begin{subfigure}[t]{0.47\linewidth}
        \centering
        \includegraphics[width=\linewidth]{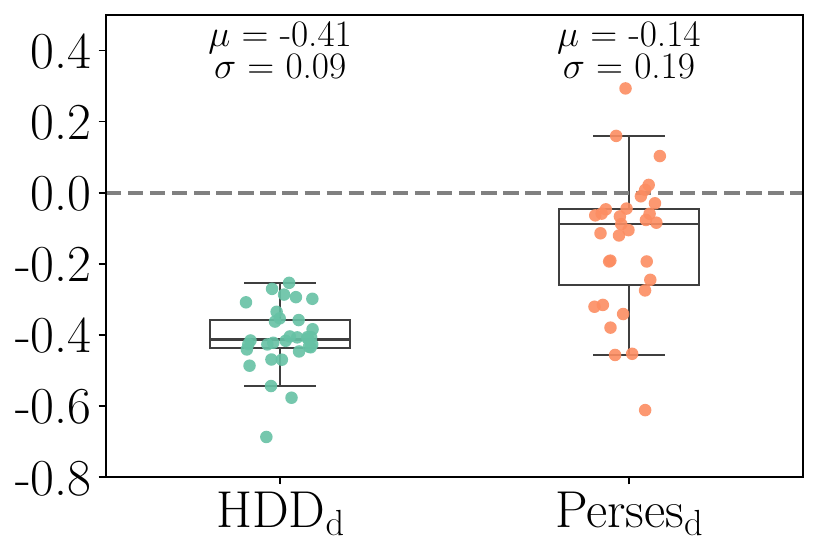}
        \caption{C Programs}
        \label{subfig:c_programs}
    \end{subfigure}
\hfil
    \begin{subfigure}[t]{0.47\linewidth}
        \centering
        \includegraphics[width=\linewidth]{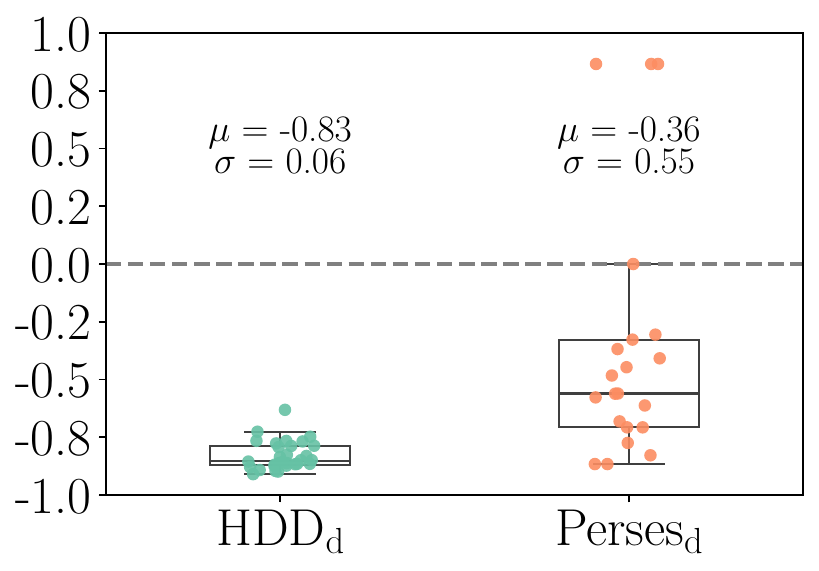}
        \caption{XML inputs}
        \label{subfig:xml_inputs}
    \end{subfigure}
    \caption{
        The Spearman correlation coefficient $\rho$ between the weights
        of elements and their probabilities being deleted in \ddmin.
        Each data point represents the mean of the $\rho$ values of all \ddmin executions on a benchmark.
    }
    \label{fig:rq1-box-plot}
\end{figure}

As shown in \cref{fig:rq1-box-plot},
overall, in each scenario of \hddDD and \persesDD,
and for both C programs and XML inputs,
our assumption is preserved. As shown in \cref{fig:rq1-box-plot},
in the four scenarios, only 5 cases of C programs and 3 cases of XML inputs
in \persesDD have $\rho$ values greater than 0,
while all other cases have $\rho$ values less or equal to 0.
Specifically, when minimizing the C programs with \hddDD and \persesDD,
the mean $\rho$ values are \spearmanMeanOfHDDDdminOnC and \spearmanMeanOfPersesDdminOnC, respectively.
For the XML inputs, the mean $\rho$ values are
\spearmanMeanOfHDDDdminOnXml and \spearmanMeanOfPersesDdminOnXml, respectively.
Although the $\rho$ values vary across different applications and benchmarks,
all are less than 0, indicating a negative correlation between the probability
of elements being deleted and their weights in \ddmin executions,
thus validating our assumption.

\finding{
    The probability of elements being deleted is negatively
    correlated with their weights in \ddmin executions
    in both \hdd and \perses, to varying degrees.
    This validation provides a solid foundation for the design of \wddmin.
}

\begin{table*}[t!]
    \centering
    \setlength{\tabcolsep}{8pt}
    \renewcommand{\arraystretch}{0.92}
    \caption{
        Results of all algorithms in \hdd and \perses on all benchmarks.
        Better results in each pair are highlighted in bold.
    }
    \label{tab:evaluation-all}
    \resizebox{0.98\textwidth}{!}{%
    \begin{tabular}{@{}crrrrrrrrrrrrrrrr@{}}
    \toprule
                                & \multicolumn{2}{c}{\hddDD}                 & \multicolumn{2}{c}{\hddWDD}                & \multicolumn{2}{c}{\persesDD}             & \multicolumn{2}{c}{\persesWDD}            & \multicolumn{2}{c}{\hddProbDD}             & \multicolumn{2}{c}{\hddWProbDD}           & \multicolumn{2}{c}{\persesProbDD}         & \multicolumn{2}{c}{\persesWProbDD}        \\ \cmidrule(){2-9} \cmidrule(){10-17}
    \multirow{-2}{*}{Benchmark} & \cellcolor[HTML]{EFEFEF}T(s)           & S(\#)         & \cellcolor[HTML]{ECF4FF}T(s)            & S(\#)           & \cellcolor[HTML]{EFEFEF}T(s)   & S(\#) & \cellcolor[HTML]{ECF4FF}T(s)           & S(\#) & \cellcolor[HTML]{EFEFEF}T(s)   & S(\#)  & \cellcolor[HTML]{ECF4FF}T(s)           & S(\#)         & \cellcolor[HTML]{EFEFEF}T(s)   & S(\#) & \cellcolor[HTML]{ECF4FF}T(s)           & S(\#)         \\ \midrule
    clang-18596                 & \cellcolor[HTML]{EFEFEF}15,664         & 452          & \cellcolor[HTML]{ECF4FF}\textbf{9,882}  & \textbf{414}   & \cellcolor[HTML]{EFEFEF}4,985          & 260          & \cellcolor[HTML]{ECF4FF}\textbf{4,885}  & 260          & \cellcolor[HTML]{EFEFEF}7,337           & 516          & \cellcolor[HTML]{ECF4FF}\textbf{6,371}  & \textbf{482}   & \cellcolor[HTML]{EFEFEF}4,966          & 261          & \cellcolor[HTML]{ECF4FF}\textbf{4,664} & \textbf{260} \\
    clang-19595                 & \cellcolor[HTML]{EFEFEF}11,332         & 285          & \cellcolor[HTML]{ECF4FF}\textbf{7,973}  & \textbf{260}   & \cellcolor[HTML]{EFEFEF}4,073          & 156          & \cellcolor[HTML]{ECF4FF}\textbf{3,977}  & 156          & \cellcolor[HTML]{EFEFEF}6,554           & 310          & \cellcolor[HTML]{ECF4FF}\textbf{5,787}  & \textbf{244}   & \cellcolor[HTML]{EFEFEF}4,311          & 156          & \cellcolor[HTML]{ECF4FF}\textbf{4,162} & 156          \\
    clang-20680                 & \cellcolor[HTML]{EFEFEF}24,517         & 377          & \cellcolor[HTML]{ECF4FF}\textbf{13,289} & \textbf{366}   & \cellcolor[HTML]{EFEFEF}19,090         & 533          & \cellcolor[HTML]{ECF4FF}\textbf{12,727} & \textbf{525} & \cellcolor[HTML]{EFEFEF}11,748          & 522          & \cellcolor[HTML]{ECF4FF}\textbf{9,153}  & \textbf{338}   & \cellcolor[HTML]{EFEFEF}20,864         & \textbf{532} & \cellcolor[HTML]{ECF4FF}\textbf{7,317} & 537          \\
    clang-21467                 & \cellcolor[HTML]{EFEFEF}28,907         & 455          & \cellcolor[HTML]{ECF4FF}\textbf{12,845} & \textbf{415}   & \cellcolor[HTML]{EFEFEF}6,240          & 177          & \cellcolor[HTML]{ECF4FF}\textbf{6,197}  & 177          & \cellcolor[HTML]{EFEFEF}7,898           & 423          & \cellcolor[HTML]{ECF4FF}\textbf{6,872}  & \textbf{310}   & \cellcolor[HTML]{EFEFEF}6,162          & \textbf{169} & \cellcolor[HTML]{ECF4FF}\textbf{5,951} & 177          \\
    clang-21582                 & \cellcolor[HTML]{EFEFEF}25,012         & 1,197        & \cellcolor[HTML]{ECF4FF}\textbf{13,257} & \textbf{999}   & \cellcolor[HTML]{EFEFEF}7,291          & 559          & \cellcolor[HTML]{ECF4FF}\textbf{7,156}  & 559          & \cellcolor[HTML]{EFEFEF}6,499           & 1,112        & \cellcolor[HTML]{ECF4FF}\textbf{4,892}  & \textbf{1,089} & \cellcolor[HTML]{EFEFEF}5,770          & 626          & \cellcolor[HTML]{ECF4FF}\textbf{5,475} & \textbf{559} \\
    clang-22337                 & \cellcolor[HTML]{EFEFEF}42,489         & \textbf{299} & \cellcolor[HTML]{ECF4FF}\textbf{16,023} & 323            & \cellcolor[HTML]{EFEFEF}2,986          & 236          & \cellcolor[HTML]{ECF4FF}\textbf{2,757}  & 236          & \cellcolor[HTML]{EFEFEF}2,176           & 338          & \cellcolor[HTML]{ECF4FF}\textbf{2,090}  & \textbf{301}   & \cellcolor[HTML]{EFEFEF}1,679          & 263          & \cellcolor[HTML]{ECF4FF}\textbf{1,648} & \textbf{250} \\
    clang-22382                 & \cellcolor[HTML]{EFEFEF}11,601         & \textbf{182} & \cellcolor[HTML]{ECF4FF}\textbf{5,332}  & 194            & \cellcolor[HTML]{EFEFEF}887            & 144          & \cellcolor[HTML]{ECF4FF}\textbf{879}    & 144          & \cellcolor[HTML]{EFEFEF}893             & \textbf{197} & \cellcolor[HTML]{ECF4FF}\textbf{791}    & 200            & \cellcolor[HTML]{EFEFEF}473            & \textbf{142} & \cellcolor[HTML]{ECF4FF}\textbf{460}   & 144          \\
    clang-22704                 & \cellcolor[HTML]{EFEFEF}49,870         & \textbf{95}  & \cellcolor[HTML]{ECF4FF}\textbf{31,873} & 97             & \cellcolor[HTML]{EFEFEF}5,627          & 78           & \cellcolor[HTML]{ECF4FF}\textbf{5,155}  & 78           & \cellcolor[HTML]{EFEFEF}6,987           & 122          & \cellcolor[HTML]{ECF4FF}\textbf{5,268}  & \textbf{90}    & \cellcolor[HTML]{EFEFEF}\textbf{2,915} & 78           & \cellcolor[HTML]{ECF4FF}5,427          & \textbf{72}  \\
    clang-23309                 & \cellcolor[HTML]{EFEFEF}54,142         & 1,085        & \cellcolor[HTML]{ECF4FF}\textbf{23,356} & \textbf{1,075} & \cellcolor[HTML]{EFEFEF}\textbf{2,995} & 475          & \cellcolor[HTML]{ECF4FF}3,266           & 475          & \cellcolor[HTML]{EFEFEF}4,962           & 1,136        & \cellcolor[HTML]{ECF4FF}\textbf{4,249}  & \textbf{1,089} & \cellcolor[HTML]{EFEFEF}1,701          & \textbf{457} & \cellcolor[HTML]{ECF4FF}\textbf{1,506} & 483          \\
    clang-23353                 & \cellcolor[HTML]{EFEFEF}61,023         & \textbf{144} & \cellcolor[HTML]{ECF4FF}\textbf{38,212} & 153            & \cellcolor[HTML]{EFEFEF}3,868          & 98           & \cellcolor[HTML]{ECF4FF}\textbf{3,005}  & 98           & \cellcolor[HTML]{EFEFEF}3,252           & \textbf{185} & \cellcolor[HTML]{ECF4FF}\textbf{2,814}  & 218            & \cellcolor[HTML]{EFEFEF}1,261          & 149          & \cellcolor[HTML]{ECF4FF}\textbf{1,240} & \textbf{98}  \\
    clang-25900                 & \cellcolor[HTML]{EFEFEF}39,481         & 478          & \cellcolor[HTML]{ECF4FF}\textbf{10,486} & \textbf{346}   & \cellcolor[HTML]{EFEFEF}2,690          & 252          & \cellcolor[HTML]{ECF4FF}\textbf{2,248}  & 252          & \cellcolor[HTML]{EFEFEF}2,312           & 518          & \cellcolor[HTML]{ECF4FF}\textbf{2,206}  & \textbf{471}   & \cellcolor[HTML]{EFEFEF}1,014          & \textbf{238} & \cellcolor[HTML]{ECF4FF}\textbf{970}   & 252          \\
    clang-26350                 & \cellcolor[HTML]{EFEFEF}72,997         & 391          & \cellcolor[HTML]{ECF4FF}\textbf{35,172} & \textbf{369}   & \cellcolor[HTML]{EFEFEF}9,494          & 189          & \cellcolor[HTML]{ECF4FF}\textbf{9,418}  & 189          & \cellcolor[HTML]{EFEFEF}12,165          & 599          & \cellcolor[HTML]{ECF4FF}\textbf{11,673} & \textbf{455}   & \cellcolor[HTML]{EFEFEF}5,395          & \textbf{232} & \cellcolor[HTML]{ECF4FF}\textbf{5,046} & 241          \\
    clang-26760                 & \cellcolor[HTML]{EFEFEF}37,413         & 321          & \cellcolor[HTML]{ECF4FF}\textbf{11,487} & \textbf{229}   & \cellcolor[HTML]{EFEFEF}4,543          & 91           & \cellcolor[HTML]{ECF4FF}\textbf{3,989}  & 91           & \cellcolor[HTML]{EFEFEF}3,678           & 306          & \cellcolor[HTML]{ECF4FF}\textbf{2,901}  & \textbf{276}   & \cellcolor[HTML]{EFEFEF}2,129          & 112          & \cellcolor[HTML]{ECF4FF}\textbf{1,922} & \textbf{90}  \\
    clang-27137                 & \cellcolor[HTML]{EFEFEF}207,071        & 582          & \cellcolor[HTML]{ECF4FF}\textbf{74,903} & \textbf{545}   & \cellcolor[HTML]{EFEFEF}14,322         & 268          & \cellcolor[HTML]{ECF4FF}\textbf{13,094} & 268          & \cellcolor[HTML]{EFEFEF}20,225          & \textbf{638} & \cellcolor[HTML]{ECF4FF}\textbf{18,724} & 661            & \cellcolor[HTML]{EFEFEF}\textbf{7,616} & \textbf{196} & \cellcolor[HTML]{ECF4FF}7,992          & 268          \\
    clang-27747                 & \cellcolor[HTML]{EFEFEF}4,468          & 299          & \cellcolor[HTML]{ECF4FF}\textbf{2,684}  & \textbf{244}   & \cellcolor[HTML]{EFEFEF}2,160          & 117          & \cellcolor[HTML]{ECF4FF}\textbf{1,464}  & 117          & \cellcolor[HTML]{EFEFEF}1,493           & 324          & \cellcolor[HTML]{ECF4FF}\textbf{1,035}  & \textbf{302}   & \cellcolor[HTML]{EFEFEF}1,187          & \textbf{137} & \cellcolor[HTML]{ECF4FF}\textbf{1,027} & 150          \\
    clang-31259                 & \cellcolor[HTML]{EFEFEF}16,977         & \textbf{556} & \cellcolor[HTML]{ECF4FF}\textbf{9,863}  & 594            & \cellcolor[HTML]{EFEFEF}3,134          & 384          & \cellcolor[HTML]{ECF4FF}\textbf{2,964}  & 384          & \cellcolor[HTML]{EFEFEF}\textbf{3,967}  & 576          & \cellcolor[HTML]{ECF4FF}5,307           & \textbf{571}   & \cellcolor[HTML]{EFEFEF}2,229          & 393          & \cellcolor[HTML]{ECF4FF}\textbf{2,200} & \textbf{384} \\
    gcc-58731                   & \cellcolor[HTML]{EFEFEF}8,369          & 431          & \cellcolor[HTML]{ECF4FF}\textbf{5,614}  & \textbf{375}   & \cellcolor[HTML]{EFEFEF}3,389          & 213          & \cellcolor[HTML]{ECF4FF}\textbf{3,325}  & 213          & \cellcolor[HTML]{EFEFEF}4,807           & 416          & \cellcolor[HTML]{ECF4FF}\textbf{3,428}  & \textbf{368}   & \cellcolor[HTML]{EFEFEF}3,112          & 237          & \cellcolor[HTML]{ECF4FF}\textbf{2,779} & \textbf{213} \\
    gcc-59903                   & \cellcolor[HTML]{EFEFEF}25,965         & \textbf{545} & \cellcolor[HTML]{ECF4FF}\textbf{12,006} & 734            & \cellcolor[HTML]{EFEFEF}4,815          & 497          & \cellcolor[HTML]{ECF4FF}\textbf{4,231}  & \textbf{382} & \cellcolor[HTML]{EFEFEF}\textbf{3,729}  & 771          & \cellcolor[HTML]{ECF4FF}5,217           & \textbf{410}   & \cellcolor[HTML]{EFEFEF}3,634          & 381          & \cellcolor[HTML]{ECF4FF}\textbf{3,020} & \textbf{300} \\
    gcc-60116                   & \cellcolor[HTML]{EFEFEF}24,708         & 1,281        & \cellcolor[HTML]{ECF4FF}\textbf{13,077} & \textbf{1,137} & \cellcolor[HTML]{EFEFEF}\textbf{3,177} & 443          & \cellcolor[HTML]{ECF4FF}3,252           & 443          & \cellcolor[HTML]{EFEFEF}5,323           & 1,245        & \cellcolor[HTML]{ECF4FF}\textbf{4,083}  & \textbf{889}   & \cellcolor[HTML]{EFEFEF}2,196          & 428          & \cellcolor[HTML]{ECF4FF}\textbf{2,092} & \textbf{404} \\
    gcc-60452                   & \cellcolor[HTML]{EFEFEF}40,082         & 491          & \cellcolor[HTML]{ECF4FF}\textbf{13,883} & \textbf{442}   & \cellcolor[HTML]{EFEFEF}2,544          & 350          & \cellcolor[HTML]{ECF4FF}\textbf{2,412}  & 350          & \cellcolor[HTML]{EFEFEF}3,559           & 824          & \cellcolor[HTML]{ECF4FF}\textbf{2,164}  & \textbf{495}   & \cellcolor[HTML]{EFEFEF}1,676          & \textbf{346} & \cellcolor[HTML]{ECF4FF}\textbf{1,656} & 350          \\
    gcc-61047                   & \cellcolor[HTML]{EFEFEF}13,086         & \textbf{495} & \cellcolor[HTML]{ECF4FF}\textbf{5,803}  & 505            & \cellcolor[HTML]{EFEFEF}1,082          & 266          & \cellcolor[HTML]{ECF4FF}\textbf{1,024}  & 266          & \cellcolor[HTML]{EFEFEF}\textbf{1,686}  & 564          & \cellcolor[HTML]{ECF4FF}2,180           & \textbf{512}   & \cellcolor[HTML]{EFEFEF}\textbf{796}   & 267          & \cellcolor[HTML]{ECF4FF}799            & \textbf{266} \\
    gcc-61383                   & \cellcolor[HTML]{EFEFEF}22,599         & 579          & \cellcolor[HTML]{ECF4FF}\textbf{10,207} & \textbf{421}   & \cellcolor[HTML]{EFEFEF}\textbf{3,070} & \textbf{271} & \cellcolor[HTML]{ECF4FF}3,187           & 274          & \cellcolor[HTML]{EFEFEF}3,524           & \textbf{509} & \cellcolor[HTML]{ECF4FF}\textbf{3,505}  & 514            & \cellcolor[HTML]{EFEFEF}2,768          & 282          & \cellcolor[HTML]{ECF4FF}\textbf{2,759} & \textbf{274} \\
    gcc-61917                   & \cellcolor[HTML]{EFEFEF}27,399         & \textbf{293} & \cellcolor[HTML]{ECF4FF}\textbf{12,531} & 311            & \cellcolor[HTML]{EFEFEF}2,414          & 142          & \cellcolor[HTML]{ECF4FF}\textbf{2,032}  & 142          & \cellcolor[HTML]{EFEFEF}2,862           & \textbf{327} & \cellcolor[HTML]{ECF4FF}\textbf{2,494}  & 300            & \cellcolor[HTML]{EFEFEF}1,845          & 145          & \cellcolor[HTML]{ECF4FF}\textbf{1,386} & \textbf{142} \\
    gcc-64990                   & \cellcolor[HTML]{EFEFEF}65,997         & \textbf{325} & \cellcolor[HTML]{ECF4FF}\textbf{36,885} & 378            & \cellcolor[HTML]{EFEFEF}5,216          & 239          & \cellcolor[HTML]{ECF4FF}\textbf{4,403}  & 239          & \cellcolor[HTML]{EFEFEF}9,047           & 601          & \cellcolor[HTML]{ECF4FF}\textbf{8,388}  & \textbf{467}   & \cellcolor[HTML]{EFEFEF}3,376          & 239          & \cellcolor[HTML]{ECF4FF}\textbf{3,342} & 239          \\
    gcc-65383                   & \cellcolor[HTML]{EFEFEF}27,974         & 246          & \cellcolor[HTML]{ECF4FF}\textbf{12,296} & \textbf{217}   & \cellcolor[HTML]{EFEFEF}1,593          & 153          & \cellcolor[HTML]{ECF4FF}\textbf{1,420}  & 153          & \cellcolor[HTML]{EFEFEF}2,273           & 281          & \cellcolor[HTML]{ECF4FF}\textbf{1,800}  & \textbf{275}   & \cellcolor[HTML]{EFEFEF}1,139          & 153          & \cellcolor[HTML]{ECF4FF}\textbf{1,113} & 153          \\
    gcc-66186                   & \cellcolor[HTML]{EFEFEF}22,124         & 605          & \cellcolor[HTML]{ECF4FF}\textbf{10,091} & \textbf{508}   & \cellcolor[HTML]{EFEFEF}2,939          & 327          & \cellcolor[HTML]{ECF4FF}\textbf{2,864}  & 327          & \cellcolor[HTML]{EFEFEF}6,688           & 591          & \cellcolor[HTML]{ECF4FF}\textbf{6,681}  & \textbf{503}   & \cellcolor[HTML]{EFEFEF}2,691          & 327          & \cellcolor[HTML]{ECF4FF}\textbf{2,603} & 327          \\
    gcc-66375                   & \cellcolor[HTML]{EFEFEF}78,050         & 1,053        & \cellcolor[HTML]{ECF4FF}\textbf{28,445} & \textbf{740}   & \cellcolor[HTML]{EFEFEF}4,114          & 440          & \cellcolor[HTML]{ECF4FF}\textbf{3,987}  & 440          & \cellcolor[HTML]{EFEFEF}\textbf{10,635} & 842          & \cellcolor[HTML]{ECF4FF}11,403          & \textbf{813}   & \cellcolor[HTML]{EFEFEF}3,345          & 440          & \cellcolor[HTML]{ECF4FF}\textbf{3,105} & 440          \\
    gcc-66412                   & \cellcolor[HTML]{EFEFEF}32,524         & 491          & \cellcolor[HTML]{ECF4FF}\textbf{13,276} & \textbf{442}   & \cellcolor[HTML]{EFEFEF}2,642          & 350          & \cellcolor[HTML]{ECF4FF}\textbf{2,421}  & 350          & \cellcolor[HTML]{EFEFEF}3,301           & 769          & \cellcolor[HTML]{ECF4FF}\textbf{2,278}  & \textbf{495}   & \cellcolor[HTML]{EFEFEF}1,794          & 350          & \cellcolor[HTML]{ECF4FF}\textbf{1,686} & 350          \\
    gcc-66691                   & \cellcolor[HTML]{EFEFEF}18,230         & 1,076        & \cellcolor[HTML]{ECF4FF}\textbf{11,607} & \textbf{1,022} & \cellcolor[HTML]{EFEFEF}3,671          & 746          & \cellcolor[HTML]{ECF4FF}\textbf{3,553}  & \textbf{602} & \cellcolor[HTML]{EFEFEF}\textbf{3,787}  & \textbf{959} & \cellcolor[HTML]{ECF4FF}5,625           & 1,039          & \cellcolor[HTML]{EFEFEF}3,256          & 689          & \cellcolor[HTML]{ECF4FF}\textbf{3,184} & \textbf{603} \\
    gcc-70127                   & \cellcolor[HTML]{EFEFEF}73,476         & 617          & \cellcolor[HTML]{ECF4FF}\textbf{30,417} & \textbf{576}   & \cellcolor[HTML]{EFEFEF}4,511          & 301          & \cellcolor[HTML]{ECF4FF}\textbf{4,279}  & 301          & \cellcolor[HTML]{EFEFEF}16,367          & 660          & \cellcolor[HTML]{ECF4FF}\textbf{11,445} & \textbf{635}   & \cellcolor[HTML]{EFEFEF}3,406          & 301          & \cellcolor[HTML]{ECF4FF}\textbf{3,234} & 301          \\
    gcc-70586                   & \cellcolor[HTML]{EFEFEF}66,155         & \textbf{792} & \cellcolor[HTML]{ECF4FF}\textbf{43,507} & 793            & \cellcolor[HTML]{EFEFEF}\textbf{6,998} & \textbf{197} & \cellcolor[HTML]{ECF4FF}7,779           & 367          & \cellcolor[HTML]{EFEFEF}13,478          & 921          & \cellcolor[HTML]{ECF4FF}\textbf{11,363} & \textbf{763}   & \cellcolor[HTML]{EFEFEF}5,812          & \textbf{168} & \cellcolor[HTML]{ECF4FF}\textbf{5,394} & 197          \\
    gcc-71626                   & \cellcolor[HTML]{EFEFEF}1,742          & 53           & \cellcolor[HTML]{ECF4FF}\textbf{409}    & 53             & \cellcolor[HTML]{EFEFEF}\textbf{50}    & 51           & \cellcolor[HTML]{ECF4FF}53              & 51           & \cellcolor[HTML]{EFEFEF}146             & 53           & \cellcolor[HTML]{ECF4FF}\textbf{104}    & 53             & \cellcolor[HTML]{EFEFEF}\textbf{45}    & 51           & \cellcolor[HTML]{ECF4FF}46             & 51           \\ \midrule
    Mean                        & \cellcolor[HTML]{EFEFEF}39,108         & 518          & \cellcolor[HTML]{ECF4FF}\textbf{18,022} & \textbf{477}   & \cellcolor[HTML]{EFEFEF}4,582          & 281          & \cellcolor[HTML]{ECF4FF}\textbf{4,169}  & \textbf{278} & \cellcolor[HTML]{EFEFEF}6,042           & 567          & \cellcolor[HTML]{ECF4FF}\textbf{5,384}  & \textbf{488}   & \cellcolor[HTML]{EFEFEF}3,455          & 280          & \cellcolor[HTML]{ECF4FF}\textbf{2,975} & \textbf{273} \\ \midrule
    xml-1                       & \cellcolor[HTML]{EFEFEF}732            & 33           & \cellcolor[HTML]{ECF4FF}\textbf{488}    & \textbf{24}    & \cellcolor[HTML]{EFEFEF}\textbf{259}   & 16           & \cellcolor[HTML]{ECF4FF}262             & 16           & \cellcolor[HTML]{EFEFEF}1,248           & 43           & \cellcolor[HTML]{ECF4FF}\textbf{528}    & \textbf{24}    & \cellcolor[HTML]{EFEFEF}425            & 16           & \cellcolor[HTML]{ECF4FF}\textbf{419}   & 16           \\
    xml-2                       & \cellcolor[HTML]{EFEFEF}1,798          & 60           & \cellcolor[HTML]{ECF4FF}\textbf{283}    & \textbf{15}    & \cellcolor[HTML]{EFEFEF}89             & 15           & \cellcolor[HTML]{ECF4FF}\textbf{75}     & 15           & \cellcolor[HTML]{EFEFEF}1,935           & 51           & \cellcolor[HTML]{ECF4FF}\textbf{283}    & \textbf{15}    & \cellcolor[HTML]{EFEFEF}139            & 15           & \cellcolor[HTML]{ECF4FF}\textbf{134}   & 15           \\
    xml-3                       & \cellcolor[HTML]{EFEFEF}765            & 36           & \cellcolor[HTML]{ECF4FF}\textbf{277}    & \textbf{15}    & \cellcolor[HTML]{EFEFEF}\textbf{81}    & 15           & \cellcolor[HTML]{ECF4FF}82              & 15           & \cellcolor[HTML]{EFEFEF}487             & 23           & \cellcolor[HTML]{ECF4FF}\textbf{236}    & \textbf{15}    & \cellcolor[HTML]{EFEFEF}\textbf{115}   & 15           & \cellcolor[HTML]{ECF4FF}128            & 15           \\
    xml-4                       & \cellcolor[HTML]{EFEFEF}2,545          & 78           & \cellcolor[HTML]{ECF4FF}\textbf{2,509}  & 78             & \cellcolor[HTML]{EFEFEF}1,159          & 13           & \cellcolor[HTML]{ECF4FF}\textbf{1,143}  & 13           & \cellcolor[HTML]{EFEFEF}2,197           & 78           & \cellcolor[HTML]{ECF4FF}\textbf{2,151}  & 78             & \cellcolor[HTML]{EFEFEF}1,234          & 13           & \cellcolor[HTML]{ECF4FF}\textbf{1,229} & 13           \\
    xml-5                       & \cellcolor[HTML]{EFEFEF}865            & 33           & \cellcolor[HTML]{ECF4FF}\textbf{242}    & \textbf{15}    & \cellcolor[HTML]{EFEFEF}\textbf{86}    & 15           & \cellcolor[HTML]{ECF4FF}89              & 15           & \cellcolor[HTML]{EFEFEF}273             & 20           & \cellcolor[HTML]{ECF4FF}\textbf{150}    & \textbf{15}    & \cellcolor[HTML]{EFEFEF}69             & 15           & \cellcolor[HTML]{ECF4FF}69             & 15           \\
    xml-6                       & \cellcolor[HTML]{EFEFEF}4,100          & 120          & \cellcolor[HTML]{ECF4FF}\textbf{4,024}  & 120            & \cellcolor[HTML]{EFEFEF}667            & 30           & \cellcolor[HTML]{ECF4FF}\textbf{659}    & 30           & \cellcolor[HTML]{EFEFEF}2,688           & 67           & \cellcolor[HTML]{ECF4FF}\textbf{1,289}  & \textbf{30}    & \cellcolor[HTML]{EFEFEF}1,145          & 30           & \cellcolor[HTML]{ECF4FF}\textbf{1,119} & 30           \\
    xml-7                       & \cellcolor[HTML]{EFEFEF}1,837          & 69           & \cellcolor[HTML]{ECF4FF}\textbf{1,787}  & 69             & \cellcolor[HTML]{EFEFEF}640            & 33           & \cellcolor[HTML]{ECF4FF}\textbf{629}    & 33           & \cellcolor[HTML]{EFEFEF}2,295           & 69           & \cellcolor[HTML]{ECF4FF}2,295           & 69             & \cellcolor[HTML]{EFEFEF}\textbf{1,062} & 33           & \cellcolor[HTML]{ECF4FF}1,080          & 33           \\
    xml-8                       & \cellcolor[HTML]{EFEFEF}4,302          & 138          & \cellcolor[HTML]{ECF4FF}\textbf{429}    & \textbf{24}    & \cellcolor[HTML]{EFEFEF}388            & 16           & \cellcolor[HTML]{ECF4FF}\textbf{378}    & 16           & \cellcolor[HTML]{EFEFEF}2,615           & 78           & \cellcolor[HTML]{ECF4FF}\textbf{531}    & \textbf{24}    & \cellcolor[HTML]{EFEFEF}\textbf{539}   & 16           & \cellcolor[HTML]{ECF4FF}546            & 16           \\
    xml-9                       & \cellcolor[HTML]{EFEFEF}\textbf{2,182} & 63           & \cellcolor[HTML]{ECF4FF}2,216           & 63             & \cellcolor[HTML]{EFEFEF}1,616          & 37           & \cellcolor[HTML]{ECF4FF}\textbf{1,591}  & 37           & \cellcolor[HTML]{EFEFEF}\textbf{1,756}  & \textbf{68}  & \cellcolor[HTML]{ECF4FF}2,272           & 90             & \cellcolor[HTML]{EFEFEF}\textbf{1,446} & 37           & \cellcolor[HTML]{ECF4FF}1,469          & 37           \\
    xml-10                      & \cellcolor[HTML]{EFEFEF}2,681          & 102          & \cellcolor[HTML]{ECF4FF}\textbf{2,651}  & 102            & \cellcolor[HTML]{EFEFEF}1,411          & 37           & \cellcolor[HTML]{ECF4FF}\textbf{1,105}  & 37           & \cellcolor[HTML]{EFEFEF}1,993           & 102          & \cellcolor[HTML]{ECF4FF}\textbf{1,937}  & 102            & \cellcolor[HTML]{EFEFEF}1,155          & 37           & \cellcolor[HTML]{ECF4FF}\textbf{999}   & 37           \\
    xml-11                      & \cellcolor[HTML]{EFEFEF}1,909          & 90           & \cellcolor[HTML]{ECF4FF}\textbf{435}    & \textbf{24}    & \cellcolor[HTML]{EFEFEF}378            & 16           & \cellcolor[HTML]{ECF4FF}\textbf{386}    & 16           & \cellcolor[HTML]{EFEFEF}759             & 37           & \cellcolor[HTML]{ECF4FF}\textbf{450}    & \textbf{24}    & \cellcolor[HTML]{EFEFEF}471            & 16           & \cellcolor[HTML]{ECF4FF}\textbf{465}   & 16           \\
    xml-12                      & \cellcolor[HTML]{EFEFEF}1,637          & 96           & \cellcolor[HTML]{ECF4FF}\textbf{1,384}  & \textbf{87}    & \cellcolor[HTML]{EFEFEF}462            & 16           & \cellcolor[HTML]{ECF4FF}\textbf{439}    & 16           & \cellcolor[HTML]{EFEFEF}2,029           & 91           & \cellcolor[HTML]{ECF4FF}\textbf{1,872}  & \textbf{87}    & \cellcolor[HTML]{EFEFEF}752            & 16           & \cellcolor[HTML]{ECF4FF}\textbf{712}   & 16           \\
    xml-13                      & \cellcolor[HTML]{EFEFEF}1,359          & 78           & \cellcolor[HTML]{ECF4FF}\textbf{1,332}  & 78             & \cellcolor[HTML]{EFEFEF}485            & 25           & \cellcolor[HTML]{ECF4FF}\textbf{483}    & 25           & \cellcolor[HTML]{EFEFEF}1,589           & 78           & \cellcolor[HTML]{ECF4FF}\textbf{1,519}  & 78             & \cellcolor[HTML]{EFEFEF}\textbf{718}   & 25           & \cellcolor[HTML]{ECF4FF}739            & 25           \\
    xml-14                      & \cellcolor[HTML]{EFEFEF}4,034          & 153          & \cellcolor[HTML]{ECF4FF}\textbf{3,955}  & 153            & \cellcolor[HTML]{EFEFEF}\textbf{1,551} & 43           & \cellcolor[HTML]{ECF4FF}1,553           & 43           & \cellcolor[HTML]{EFEFEF}4,882           & 160          & \cellcolor[HTML]{ECF4FF}\textbf{4,716}  & \textbf{153}   & \cellcolor[HTML]{EFEFEF}\textbf{2,267} & 43           & \cellcolor[HTML]{ECF4FF}2,292          & 43           \\
    xml-15                      & \cellcolor[HTML]{EFEFEF}2,333          & 108          & \cellcolor[HTML]{ECF4FF}\textbf{1,012}  & \textbf{51}    & \cellcolor[HTML]{EFEFEF}545            & 16           & \cellcolor[HTML]{ECF4FF}\textbf{534}    & 16           & \cellcolor[HTML]{EFEFEF}1,581           & 85           & \cellcolor[HTML]{ECF4FF}\textbf{993}    & \textbf{51}    & \cellcolor[HTML]{EFEFEF}621            & 16           & \cellcolor[HTML]{ECF4FF}\textbf{620}   & 16           \\
    xml-16                      & \cellcolor[HTML]{EFEFEF}1,738          & 60           & \cellcolor[HTML]{ECF4FF}\textbf{513}    & \textbf{24}    & \cellcolor[HTML]{EFEFEF}437            & 16           & \cellcolor[HTML]{ECF4FF}\textbf{430}    & 16           & \cellcolor[HTML]{EFEFEF}1,157           & 38           & \cellcolor[HTML]{ECF4FF}\textbf{532}    & \textbf{24}    & \cellcolor[HTML]{EFEFEF}\textbf{579}   & 16           & \cellcolor[HTML]{ECF4FF}602            & 16           \\
    xml-17                      & \cellcolor[HTML]{EFEFEF}2,519          & 87           & \cellcolor[HTML]{ECF4FF}\textbf{292}    & \textbf{15}    & \cellcolor[HTML]{EFEFEF}101            & 15           & \cellcolor[HTML]{ECF4FF}\textbf{97}     & 15           & \cellcolor[HTML]{EFEFEF}1,458           & 58           & \cellcolor[HTML]{ECF4FF}\textbf{212}    & \textbf{15}    & \cellcolor[HTML]{EFEFEF}\textbf{108}   & 15           & \cellcolor[HTML]{ECF4FF}124            & 15           \\
    xml-18                      & \cellcolor[HTML]{EFEFEF}785            & 39           & \cellcolor[HTML]{ECF4FF}\textbf{761}    & 39             & \cellcolor[HTML]{EFEFEF}433            & 16           & \cellcolor[HTML]{ECF4FF}\textbf{423}    & 16           & \cellcolor[HTML]{EFEFEF}1,496           & 50           & \cellcolor[HTML]{ECF4FF}\textbf{916}    & \textbf{39}    & \cellcolor[HTML]{EFEFEF}569            & 16           & \cellcolor[HTML]{ECF4FF}\textbf{549}   & 16           \\
    xml-19                      & \cellcolor[HTML]{EFEFEF}\textbf{1,512} & 54           & \cellcolor[HTML]{ECF4FF}1,539           & 54             & \cellcolor[HTML]{EFEFEF}\textbf{628}   & 36           & \cellcolor[HTML]{ECF4FF}632             & 36           & \cellcolor[HTML]{EFEFEF}1,858           & 60           & \cellcolor[HTML]{ECF4FF}\textbf{1,807}  & \textbf{54}    & \cellcolor[HTML]{EFEFEF}\textbf{937}   & 36           & \cellcolor[HTML]{ECF4FF}968            & 36           \\
    xml-20                      & \cellcolor[HTML]{EFEFEF}2,492          & 99           & \cellcolor[HTML]{ECF4FF}\textbf{2,380}  & 99             & \cellcolor[HTML]{EFEFEF}2,117          & 64           & \cellcolor[HTML]{ECF4FF}\textbf{2,108}  & 64           & \cellcolor[HTML]{EFEFEF}\textbf{3,120}  & 99           & \cellcolor[HTML]{ECF4FF}3,209           & 99             & \cellcolor[HTML]{EFEFEF}3,502          & 64           & \cellcolor[HTML]{ECF4FF}\textbf{3,393} & 64           \\
    xml-21                      & \cellcolor[HTML]{EFEFEF}3,362          & 93           & \cellcolor[HTML]{ECF4FF}\textbf{3,256}  & \textbf{90}    & \cellcolor[HTML]{EFEFEF}1,550          & 46           & \cellcolor[HTML]{ECF4FF}\textbf{1,463}  & 46           & \cellcolor[HTML]{EFEFEF}4,244           & 91           & \cellcolor[HTML]{ECF4FF}\textbf{4,192}  & \textbf{90}    & \cellcolor[HTML]{EFEFEF}\textbf{2,480} & 46           & \cellcolor[HTML]{ECF4FF}2,549          & 46           \\
    xml-22                      & \cellcolor[HTML]{EFEFEF}1,454          & 61           & \cellcolor[HTML]{ECF4FF}\textbf{1,402}  & 61             & \cellcolor[HTML]{EFEFEF}346            & 24           & \cellcolor[HTML]{ECF4FF}\textbf{335}    & 24           & \cellcolor[HTML]{EFEFEF}1,633           & 61           & \cellcolor[HTML]{ECF4FF}\textbf{1,467}  & 61             & \cellcolor[HTML]{EFEFEF}480            & 24           & \cellcolor[HTML]{ECF4FF}\textbf{420}   & 24           \\
    xml-23                      & \cellcolor[HTML]{EFEFEF}7,317          & 189          & \cellcolor[HTML]{ECF4FF}\textbf{5,905}  & \textbf{165}   & \cellcolor[HTML]{EFEFEF}2,332          & 54           & \cellcolor[HTML]{ECF4FF}\textbf{2,183}  & \textbf{51}  & \cellcolor[HTML]{EFEFEF}\textbf{6,725}  & \textbf{152} & \cellcolor[HTML]{ECF4FF}6,727           & 165            & \cellcolor[HTML]{EFEFEF}\textbf{3,184} & \textbf{47}  & \cellcolor[HTML]{ECF4FF}3,432          & 51           \\
    xml-24                      & \cellcolor[HTML]{EFEFEF}\textbf{2,269} & 96           & \cellcolor[HTML]{ECF4FF}2,365           & 96             & \cellcolor[HTML]{EFEFEF}2,092          & 70           & \cellcolor[HTML]{ECF4FF}\textbf{2,055}  & 70           & \cellcolor[HTML]{EFEFEF}2,896           & 96           & \cellcolor[HTML]{ECF4FF}\textbf{2,803}  & 96             & \cellcolor[HTML]{EFEFEF}3,421          & 70           & \cellcolor[HTML]{ECF4FF}\textbf{3,196} & 70           \\
    xml-25                      & \cellcolor[HTML]{EFEFEF}\textbf{3,114} & 135          & \cellcolor[HTML]{ECF4FF}3,696           & \textbf{132}   & \cellcolor[HTML]{EFEFEF}1,956          & 55           & \cellcolor[HTML]{ECF4FF}\textbf{1,924}  & 55           & \cellcolor[HTML]{EFEFEF}2,672           & 139          & \cellcolor[HTML]{ECF4FF}\textbf{2,042}  & \textbf{135}   & \cellcolor[HTML]{EFEFEF}1,745          & 55           & \cellcolor[HTML]{ECF4FF}\textbf{1,710} & 55           \\
    xml-26                      & \cellcolor[HTML]{EFEFEF}6,238          & 126          & \cellcolor[HTML]{ECF4FF}\textbf{6,166}  & 126            & \cellcolor[HTML]{EFEFEF}\textbf{2,688} & 64           & \cellcolor[HTML]{ECF4FF}2,762           & 64           & \cellcolor[HTML]{EFEFEF}6,373           & 128          & \cellcolor[HTML]{ECF4FF}\textbf{6,268}  & 138            & \cellcolor[HTML]{EFEFEF}4,278          & 64           & \cellcolor[HTML]{ECF4FF}\textbf{4,179} & 64           \\
    xml-27                      & \cellcolor[HTML]{EFEFEF}8,955          & 195          & \cellcolor[HTML]{ECF4FF}\textbf{8,769}  & 195            & \cellcolor[HTML]{EFEFEF}\textbf{4,382} & 102          & \cellcolor[HTML]{ECF4FF}4,442           & 102          & \cellcolor[HTML]{EFEFEF}9,340           & 188          & \cellcolor[HTML]{ECF4FF}\textbf{8,529}  & \textbf{186}   & \cellcolor[HTML]{EFEFEF}7,135          & 102          & \cellcolor[HTML]{ECF4FF}\textbf{7,030} & 102          \\
    xml-28                      & \cellcolor[HTML]{EFEFEF}7,690          & 159          & \cellcolor[HTML]{ECF4FF}\textbf{7,333}  & 159            & \cellcolor[HTML]{EFEFEF}\textbf{3,811} & 97           & \cellcolor[HTML]{ECF4FF}3,820           & 97           & \cellcolor[HTML]{EFEFEF}7,818           & 159          & \cellcolor[HTML]{ECF4FF}\textbf{7,203}  & 159            & \cellcolor[HTML]{EFEFEF}5,928          & 97           & \cellcolor[HTML]{ECF4FF}\textbf{5,825} & 97           \\
    xml-29                      & \cellcolor[HTML]{EFEFEF}5,860          & 147          & \cellcolor[HTML]{ECF4FF}\textbf{5,173}  & \textbf{138}   & \cellcolor[HTML]{EFEFEF}1,906          & 48           & \cellcolor[HTML]{ECF4FF}\textbf{1,861}  & 48           & \cellcolor[HTML]{EFEFEF}3,810           & 142          & \cellcolor[HTML]{ECF4FF}\textbf{3,515}  & \textbf{138}   & \cellcolor[HTML]{EFEFEF}1,819          & 48           & \cellcolor[HTML]{ECF4FF}\textbf{1,797} & 48           \\
    xml-30                      & \cellcolor[HTML]{EFEFEF}6,190          & 147          & \cellcolor[HTML]{ECF4FF}\textbf{6,054}  & 147            & \cellcolor[HTML]{EFEFEF}\textbf{4,258} & 78           & \cellcolor[HTML]{ECF4FF}4,261           & 78           & \cellcolor[HTML]{EFEFEF}7,197           & 147          & \cellcolor[HTML]{ECF4FF}\textbf{6,583}  & 147            & \cellcolor[HTML]{EFEFEF}6,807          & 78           & \cellcolor[HTML]{ECF4FF}\textbf{6,591} & 78           \\ \midrule
    Mean                        & \cellcolor[HTML]{EFEFEF}3,152          & 98           & \cellcolor[HTML]{ECF4FF}\textbf{2,621}  & \textbf{82}    & \cellcolor[HTML]{EFEFEF}1,295          & 38           & \cellcolor[HTML]{ECF4FF}\textbf{1,273}  & 38           & \cellcolor[HTML]{EFEFEF}3,004           & 89           & \cellcolor[HTML]{ECF4FF}\textbf{2,574}  & \textbf{80}    & \cellcolor[HTML]{EFEFEF}1,838          & 37           & \cellcolor[HTML]{ECF4FF}\textbf{1,813} & 37           \\ \bottomrule
    \end{tabular}%
    }
    \end{table*}

\subsection{\wddmin \vs \ddmin}
\label{subsec:rq2}
For this question, we compare the performance
of \hddWDD and \persesWDD with \hddDD and \persesDD, respectively.
The detailed results are shown in \cref{tab:evaluation-all}.

\subsubsection{Effectiveness}
\label{subsubsec:rq2-effectiveness}
Overall, \wddmin is more effective than \ddmin in both \hdd and \perses.
On average, \hddWDD generates \hddWddSizeOnC and \hddWddSizeOnXML smaller results
than \hddDD on C and XML benchmarks, respectively,
with a p-value of \pValHddWddSize overall.
In \perses,
the results of \persesWDD are \persesWddSizeOnC and \persesWddSizeOnXML
smaller than those of \persesDD on C and XML benchmarks, respectively,
with a p-value of \pValPersesWddSize overall.

Notably,
while the above results demonstrate the superior effectiveness of \wddmin over \ddmin,
the improvement of \wddmin over \ddmin in \perses is not as significant as that in \hdd.
Given the different design of \hdd and \perses, this result is expected.
Unlike \hdd that fully relies on \ddmin to perform tree node deletion,
\perses customizes different deletion strategies for different types of nodes.
In \perses, \ddmin is only used to minimize the list of nodes under a quantified node~\cite{sun2018perses}.
That is to say, compared to \hdd, the deletion operations performed by \ddmin (or \wddmin)
constitute a smaller proportion of the total operations in \perses.
Therefore, improvements to the effectiveness of \ddmin have a relatively
moderate impact on the overall effectiveness of \perses.
Moreover, the nodes under a quantified node in \perses are
syntactically independent from each other~\cite{sun2018perses}, making the minimization
less challenging. Thus, \ddmin can generate results comparable to \wddmin.

\subsubsection{Efficiency}
\label{subsubsec:rq2-efficiency}
We first evaluate efficiency with processing time, for which \wddmin outperforms \ddmin
in both \hdd and \perses. On average, \hddWDD takes \hddWddTimeOnC and
\hddWddTimeOnXML less time than \hddDD to finish minimizing
the C programs and XML inputs, respectively,
with a p-value of \pValHddWddTime overall.
Similarly, \persesWDD reduces
the processing time of \persesDD by \persesWddTimeOnC and \persesWddTimeOnXML
on C and XML benchmarks, respectively,
with a p-value of \pValPersesWddTime overall.
Furthermore, considering the number of tokens deleted per second (referred as \tokensPerSecond) as an additional metric,
while \hddDD and \persesDD deletes \hddDdminAverageTokensPerSecond
and \persesDdminAverageTokensPerSecond \tokensPerSecond, respectively,
\hddWDD and \persesWDD deletes \hddWddAverageTokensPerSecond
and \persesWddAverageTokensPerSecond \tokensPerSecond,
which are \hddWddTokensPerSecondRatio and \persesWddTokensPerSecondRatio
more than those of \hddDD and \persesDD, respectively.
These results strongly indicate that \wddmin is more efficient than \ddmin.

Similar with the improvement of effectiveness,
while \wddmin consistently achieves higher efficiency than \ddmin
in both \hdd and \perses, the improvement is more significant in \hdd than
in \perses. The reason is the same as that explained in \cref{subsubsec:rq2-effectiveness} for effectiveness.
Moreover, the high efficiency of \wddmin is based on the assumption that the probability
of elements being deleted is negatively correlated with their weights,
which is validated in \cref{subsec:rq1}.
In fact, the degree of this correlation can affect the efficiency of \wddmin.
As shown in \cref{fig:rq1-box-plot}, the $\rho$ values of \persesDD are generally larger than those of \hddDD,
indicating a weaker negative correlation. Thus, the improvement of \wddmin over \ddmin
in \perses is not as significant as that in \hdd.

\finding{
    \wddmin outperforms \ddmin in both effectiveness and efficiency in \hdd,
    by generating \hddWddAverageSize smaller results
    in \hddWddAverageTime less time on average.
    In \perses, \wddmin exceeds \ddmin in efficiecny by taking \persesWddAverageTime
    less time on average to generate the comparable results.
}

\subsection{RQ3: \wprobdd \vs \probdd}

For this research question, we compare the performance of \hddWProbDD
and \persesWProbDD using \hddProbDD and \persesProbDD as baselines, respectively.
The minimization process of \probdd contains
nondeterminism
since it may randomly select elements
when their probabilities are the same.
To mitigate the impact of such nondeterminism, we repeat each experiment for 5 times
and report the average results. \wprobdd largely eliminates the randomness of
\probdd by considering the weights of elements.
The detailed results are shown in \cref{tab:evaluation-all}.

\myparagraph{Effectiveness}
Overall, \wprobdd is more effective than \probdd by generating smaller results.
On average, \hddWProbDD generates \hddWprobddSizeOnC and \hddWprobddSizeOnXML
smaller results than \hddProbDD for the C programs and XML inputs, respectively,
with a p-value of \pValHddWprobddSize.
Besides, \persesWProbDD generates \persesWprobddSizeOnC smaller and
\persesWprobddSizeOnXML larger results than \persesProbDD for the C
programs and XML inputs, respectively,
with a p-value of \pValPersesWprobddSize.
There is no significant difference
between the results of \persesWProbDD and \persesProbDD.
In fact, \persesWProbDD generates
\persesWprobddSameResultCount same results as \persesProbDD out of \allBenchmarksSize benchmarks,
of which, \persesWprobddSameResultCountOnXML are from the XML inputs,
because of the same reason
explained in \cref{subsubsec:rq2-efficiency}.
Especially, for the XML inputs, only 11.9\% property tests performed by \persesWProbDD
are from \wprobdd,  indicating
that the effectiveness of \persesWProbDD is largely
determined by the inner deletion strategies of \perses, instead of \wprobdd.

\myparagraph{Efficiency}
\wprobdd achieves higher efficiency than \probdd in both \hdd and \perses.
We first evaluate the efficiency of \wprobdd with processing time.
On average, \hddWProbDD shortens the processing time of \hddProbDD by
\hddWprobddTimeOnC and \hddWprobddTimeOnXML for the C programs and XML inputs, respectively,
with a p-value of \pValHddWprobddTime overall.
Similarly, \persesWProbDD reduces the processing time of \persesProbDD by
\persesWprobddTimeOnC and \persesWprobddTimeOnXML on each benchmark suite, respectively,
with a p-value of \pValPersesWprobddTime overall.
Moreover, in terms of the number of tokens deleted per second as an additional metric,
while \hddProbDD and \persesProbDD deletes \hddProbddAverageTokensPerSecond
and \persesProbddAverageTokensPerSecond \tokensPerSecond,
\hddWProbDD and \persesWProbDD deletes \hddWprobddAverageTokensPerSecond
and \persesWprobddAverageTokensPerSecond \tokensPerSecond,
which are \hddWprobddTokensPerSecondRatio and \persesWprobddTokensPerSecondRatio
more than those of \hddProbDD and \persesProbDD, respectively.

\finding{
    \wprobdd outperforms \probdd in both effectiveness and efficiency,
    by making \hdd and \perses
    produce \hddWprobddAverageSize and \persesWprobddAverageSize smaller results
    in \hddWprobddAverageTime and \persesWprobddAverageTime less time
    on average, respectively.
}

\section{Discussion}
\label{sec:discussion}
\subsection{Alternative Weight Assignment}

In our implementation of \wdd in \ddmin and \probdd in this paper, we utilize
the number of tokens of each element as the weight.
Although this assignment strategy is not 100\% accurate,
it achieves high efficiency and feasibility as it is \emph{static,
lightweight and generalizable}. Other weight
assignment could also be considered, such as a dynamic weight assignment strategy based
on runtime information, including factors like memory usage, IO operations, or
execution time. However, such a dynamic weight assignment strategy may introduce additional
overhead, potentially hindering the performance of minimization. Furthermore,
runtime profiling techniques are typically language-specific, which may limit the
generalizability of \wdd.
Overcoming these challenges and exploring the potential of dynamic \wdd
for language-specific minimization techniques presents an interesting direction for future work.

\subsection{Limitations}

The primary limitation of \wdd is its applicability mainly to tree-structured inputs,
where it is most effective when the weights (\ie, token counts) of elements vary significantly.
When the test inputs cannot be represented in a tree structure (e.g., random strings),
while the concept of weight still exists, token count may not serve as an appropriate
weight representation. Additionally, if the tree representation of the test input is
highly balanced, \wdd may offer only marginal improvement over traditional delta debugging methods.
Nevertheless, given the widespread use of tree-based minimization techniques and the
typically unbalanced nature of trees in real-world inputs, \wdd remains essential
for enhancing the performance of test input minimization in practical scenarios.

\subsection{Threats to Validity}
\label{sec:threats}
\subsubsection{Threats to Internal Validity}
The primary internal threat arises from the implementation of the evaluated
techniques, including \wddmin, \wprobdd, and their respective baselines,
as well as \hdd and \perses.
To mitigate this threat, we rigorously reproduced the the baseline
techniques based on their descriptions in the original papers,
and wrote multiple test cases to ensure the algorithms functioned as expected.
Additionally, all authors of this paper participated in a thorough code review
of the implementation. Prior to evaluating the full set of benchmarks,
we randomly selected several cases, ran our algorithms on them, and manually
verified the detailed results to confirm the accuracy of our implementations.
We have also made our implementations publicly available for replication and
facilitating further research.

\subsubsection{Threats to External Validity}
A key threat to external validity is the generalizability of \wdd across different
input formats or languages.
Although WDD is designed to apply to all tree-structured inputs, variations in the
tree characteristics of different inputs may impact its performance.
To mitigate this threat, we evaluated WDD on two types of benchmarks: C and XML.
The C benchmarks represent traditional programming languages, while the XML files
represent structured inputs that are highly hierarchical but not programs.
Our evaluation results demonstrate the superior performance of WDD across these
diverse formats. To further address this threat, our future work
includes expanding the evaluation of WDD to a broader range of benchmarks.

\section{Related Work}
\label{sec:related_work}
We introduce two lines of related  work.

\myparagraph{Test Input Minimization}
\deltadebugging~\cite{zeller2002simplifying} is the first
systematic study that enlightens the research of test input
minimization. It introduced an minimizing algorithm
named \ddmin to minimize failure-inducing test inputs, which has been
described in \cref{subsubsec:ddmin}.
While \ddmin is effective, its efficiency is not satisfactory
as it follows a predefined pattern to partition and delete elements,
overlooking the information of existing tests. To fix this issue,
Wang et al.~\cite{wang2021probabilistic} proposed \probdd.
As explained in \cref{subsec:probdd}, \probdd leverages
a probabilistic model to guide the minimization process.
However, both \ddmin and \probdd overlook the different sizes of
elements in the list, leading to suboptimal performance.
Contrastively, our approaches, \wddmin and \wprobdd, successfully
distinguish different elements with their weights, and make more
rationale partitioning decisions with considering weights,
which significantly
improves the performance of prior delta debugging algorithms.
In practice, rather than being used directly to minimize test inputs,
delta debugging algorithms are often integrated into
tree-based minimization techniques for better performance. Two
representative techniques are \hdd~\cite{misherghi2006hdd}
and \perses~\cite{sun2018perses}, which are chosen for our evaluation.
\hdd and \perses apply delta debugging algorithms to minimize
the list of nodes in the tree. Thus their performance can be
further improved by equipping our new delta debugging algorithms.

\myparagraph{Program Reduction}
Program reduction is a special case of test input minimization, where the input
is a program. Since normally a program can be parsed into a syntax tree, tree-based
test input minimization techniques, \eg, \hdd and \perses, can be directly
applied to program reduction. Moreover, Xu \etal proposed \vulcan~\cite{xu2023pushing},
which pushes the limit of 1-minimality by performing predefined program transformations.
They further developed T-Rec~\cite{xu2024t}, a fine-grained language-agnostic program reduction
technique guided by lexical syntax. T-rec is demonstrated to not only achieve smaller
minimization results than \vulcan, but also aids in deduplicating bug-triggering test inputs.
Additionally, Zhang \etal proposed LPR~\cite{zhang2024lpr}, the first language-agnostic
program reducer boosted by large language models.
Furthermore, some program reduction techniques are specifically
designed for certain languages. For example, \creduce~\cite{regehr2012test}
is specifically designed for reducing C/C++ programs. It incorporates various
semantic-specific transformations to effectively minimize C/C++ programs.
\jreduce~\cite{kalhauge2019binary,kalhauge2021logical},
\ddsmt~\cite{niemetz2013ddsmt} and
\jsdelta~\cite{jsdelta} are specifically designed for reducing
Java bytecode, SMT-LIBv2 inputs, and JavaScript programs, respectively.
Herfert \etal propose the Generalized Tree Reduction (GTR) technique
which minimizes programs with a series of language-specific transformations generated by learning
from a corpus of example data~\cite{herfert2017automatically}.
While these approaches are designed for specific languages,
some of them, such as \ddsmt, apply delta debugging under the hood. To this end,
introducing our novel concept of \wdd to these tools to further improve their performance
is a promising direction for future work.

\section{Conclusion}
\label{sec:conclusion}
This paper introduces \weighteddeltadebugging (\wdd),
a novel concept that incorporates the weight of elements into delta debugging.
The key insight of \wdd is to assign each element in the input list a weight,
and distinguish different elements based on their weights during partitioning.
We realize the concept of \wdd in two representative delta debugging algorithms,
\ddmin and \probdd, and propose \wddmin and \wprobdd, respectively.
The extensive evaluation on \allBenchmarksSize benchmarks
demonstrates the superior performance of \wddmin and \wprobdd,
in both effectiveness and efficiency, highlighting the significance
of \wdd in optimizing delta debugging algorithms.
We firmly believe that \wdd opens up a new dimension to improve
test input minimization techniques.

\section*{Acknowledgments}
We thank all the anonymous reviewers in ICSE'25 for their insightful
feedback and comments.
This research is partially supported by
the Natural Sciences and Engineering Research Council of Canada
(NSERC) through the
Discovery Grant, a project under WHJIL,
and CFI-JELF Project \#40736.

\bibliographystyle{IEEEtran}
\bibliography{acmart}

\end{document}